\documentclass[prl,twocolumn,preprintnumbers,superscriptaddress,amsmath,amssymb]{revtex4-1}
\usepackage{graphicx}
\usepackage{subfigure}
\usepackage{mathrsfs}
\usepackage{amsfonts}
\usepackage{times}
\usepackage{amsmath}
\usepackage{leftidx}
\usepackage{tikz}
\usepackage{tikz-network}
\usepackage{color}
\usepackage[colorlinks,linkcolor=blue,citecolor=blue]{hyperref}

\newcommand{\ind}{\operatorname{ind}}
\newcommand{\Tr}{\operatorname{Tr}}

\usepackage{braket}
\usepackage{mathtools}

\begin{document}
\title{Topological lower bound on quantum chaos by entanglement growth}
\author{Zongping Gong}
\affiliation{Max-Planck-Institut f\"ur Quantenoptik, Hans-Kopfermann-Stra{\ss}e 1, D-85748 Garching, Germany}
\affiliation{Munich Center for Quantum Science and Technology, Schellingstra{\ss}e 4, 80799 M\"unchen, Germany}
\author{Lorenzo Piroli}
\affiliation{Max-Planck-Institut f\"ur Quantenoptik, Hans-Kopfermann-Stra{\ss}e 1, D-85748 Garching, Germany}
\affiliation{Munich Center for Quantum Science and Technology, Schellingstra{\ss}e 4, 80799 M\"unchen, Germany}
\author{J. Ignacio Cirac}
\affiliation{Max-Planck-Institut f\"ur Quantenoptik, Hans-Kopfermann-Stra{\ss}e 1, D-85748 Garching, Germany}
\affiliation{Munich Center for Quantum Science and Technology, Schellingstra{\ss}e 4, 80799 M\"unchen, Germany}
\date{\today}

\begin{abstract}
A fundamental result in modern quantum chaos theory is the Maldacena-Shenker-Stanford upper bound on the growth of out-of-time-order correlators, 
whose infinite-temperature limit is related to the operator-space entanglement entropy of the evolution operator. Here we show that, for one-dimensional quantum cellular automata (QCA), there exists a lower bound on quantum chaos quantified by such entanglement entropy. This lower bound is equal to twice the index of the QCA, which is a topological invariant that measures the chirality of information flow, and holds for all the R\'enyi entropies, with its strongest R\'enyi-$\infty$ version being tight. The rigorous bound rules out the possibility of any sublinear entanglement growth behavior, showing in  particular that many-body localization is forbidden for unitary evolutions displaying nonzero index. Since the R\'enyi entropy is measurable, our findings have direct experimental relevance. Our result is robust  
against exponential tails which naturally appear in quantum dynamics generated by local Hamiltonians.
\end{abstract}
\maketitle

\begin{figure}
	\begin{center}
		\includegraphics[width=8.5cm, clip]{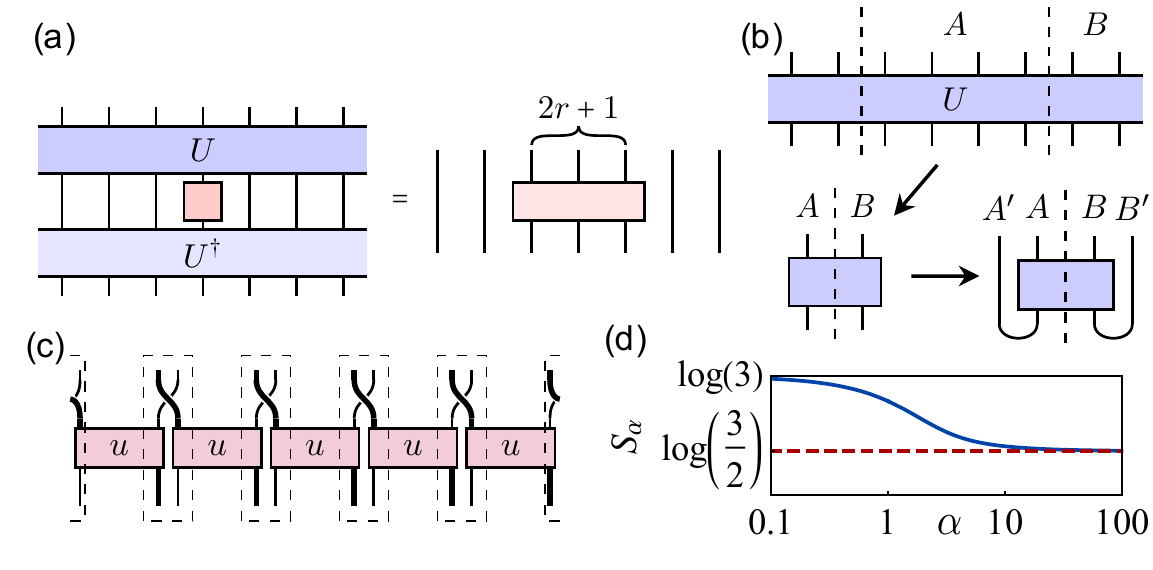}
	\end{center}
	\caption{(a) In 1D, a range-$r$ QCA is defined as a unitary $U$ which transforms an on-site operator supported on the $j$th site into an operator supported on the interval $[j-r,j+r]$. (b) Bipartition $A\sqcup B$ and vectorization of $U$ into the CJS $|U\rangle$ on a doubled Hilbert space [cf. Eq.~(\ref{CJS})]. The operator entanglement of $U$ with respect to $A\sqcup B$ is defined as the entanglement entropy of $|U\rangle$ with respect to the bipartition $AA'\sqcup B B'$, where we denote by $A'$, $B'$  the ancillary qudits associated with $A$, $B$ respectively. (c) AKLT-like range-1 QCA with local Hilbert space $\mathbb{C}^{pq}$ (dashed rectangles) ($p$ and $q$ are coprime) and $\ind=\ln(p/q)$ generated by disjoint unitaries $u:\mathbb{C}^p\otimes \mathbb{C}^q\to\mathbb{C}^q\otimes \mathbb{C}^p$. Here the thin and thick legs correspond to $\mathbb{C}^p$ and $\mathbb{C}^q$, respectively. (d) Operator entanglement R\'enyi  entropy $S_\alpha$ of a single $u$ in (c) with $p=2$ and $q=3$ approaches $|\ind| =\log(3/2)$ when $\alpha\to\infty$, implying the saturation of Eq.~(\ref{eq:main_result}) for $S_\infty$.}
	\label{fig1}
\end{figure}

\emph{Introduction.---} The principles of causality and conservation of quantum information impose strong constraints on the evolution of quantum many-body systems. In the simplest setting, where space and time are discrete and causality is ``strict'', the latter can be described by quantum cellular automata (QCA)~\cite{schumacher2004reversible,farrelly2019review,arrighi2019overview}, cf. Fig.~\ref{fig1}(a) for an illustration. Despite seemingly crude approximations for realistic many-body dynamics, they provide useful models to study different aspects of non-equilibrium physics. For instance, local quantum circuits, a subclass of QCA, recently received significant attention in connection to questions related to quantum chaos and information scrambling~\cite{nahum2017quantum,nahum2018operator,vonKeyserlingk2018operator,Tibor2018,Khemani2018,Christoph2018,Chan2018,Chan2018b,Friedman2019,Bruno2019,Bruno2020,Swingle2020,Claeys2020Maximum}.

In the past decade, much progress has been made in the characterization of QCA~\cite{arrighi2011unitarity,haah2018nontrivial,haah2019clifford,freedman2020classification,piroli2020quantum}, with comprehensive and elegant results obtained in one dimension (1D)~\cite{gross2012index,cirac2017matrix,sahinoglu2018matrix,gong2020classification,piroli2020fermionic}. In particular, it was first found in Ref.~\cite{gross2012index} that QCA are characterized by a topological index, sometimes called GNVW after the authors, which measures the amount of quantum information flowing through the system. Besides its fundamental interest, this result turned out to have practical implications, allowing, for instance, for a classification of 2D Floquet phases exhibiting bulk many-body localization (MBL)~\cite{Else2016,
Po2016,Potter2017,Roy2017,duschatko2018Tracking,fidkowski2019interacting,fidkowski2019interacting,zhang2020classification,Liu2020}.

In light of the intuitive picture in terms of flow of quantum information, it is natural to ask whether there exist strict relations between the index and other aspects of  the unitary dynamics, related, for instance, to information scrambling as probed by out-of-time-ordered correlators (OTOCs)~\cite{shenker2014multiple,shenker2014black,roberts2015localized}. The main difficulty to answer this question lies perhaps in the original definition of the index~\cite{gross2012index}, which was given in terms of the rank of certain operator algebras, lacking an immediate physical interpretation.

In this work, we prove that there is an equivalent way to express the index in terms of the entanglement of the ``vectorized" evolution operator [cf. Fig.~\ref{fig1}(b)], which is often called operator-space entanglement entropy~\cite{zanardi2001entanglement} . This quantity can be formulated in terms of any R\'enyi entropy, is computed locally, and closely reflects the intuitive interpretation of the index in terms of quantum-information flow. Inspired by this definition, we derive our main result
\begin{equation}
	\label{eq:main_result}
S_{\alpha}\geq 2|\ind|\,,
\end{equation}
where $S_{\alpha}$ is the R\'enyi entropy of order $\alpha$ of the evolution operator. This bound holds for any $\alpha\in[0,\infty]$. In particular, based on the known relation between $S_{2}$ and the average of OTOCs~\cite{hosur2016chaos}, we will interpret this result as a lower bound on quantum chaos. As an important application,  Eq.~\eqref{eq:main_result} allows us to establish rigorously that any sublinear entropy growth behavior in 1D, including MBL, is not compatible with a non-vanishing index. Thanks to the experimental accessibility to the R\'enyi entropy, our results should be observable in current quantum simulation experiments. 

\begin{figure}
\begin{center}
       \includegraphics[width=8.5cm, clip]{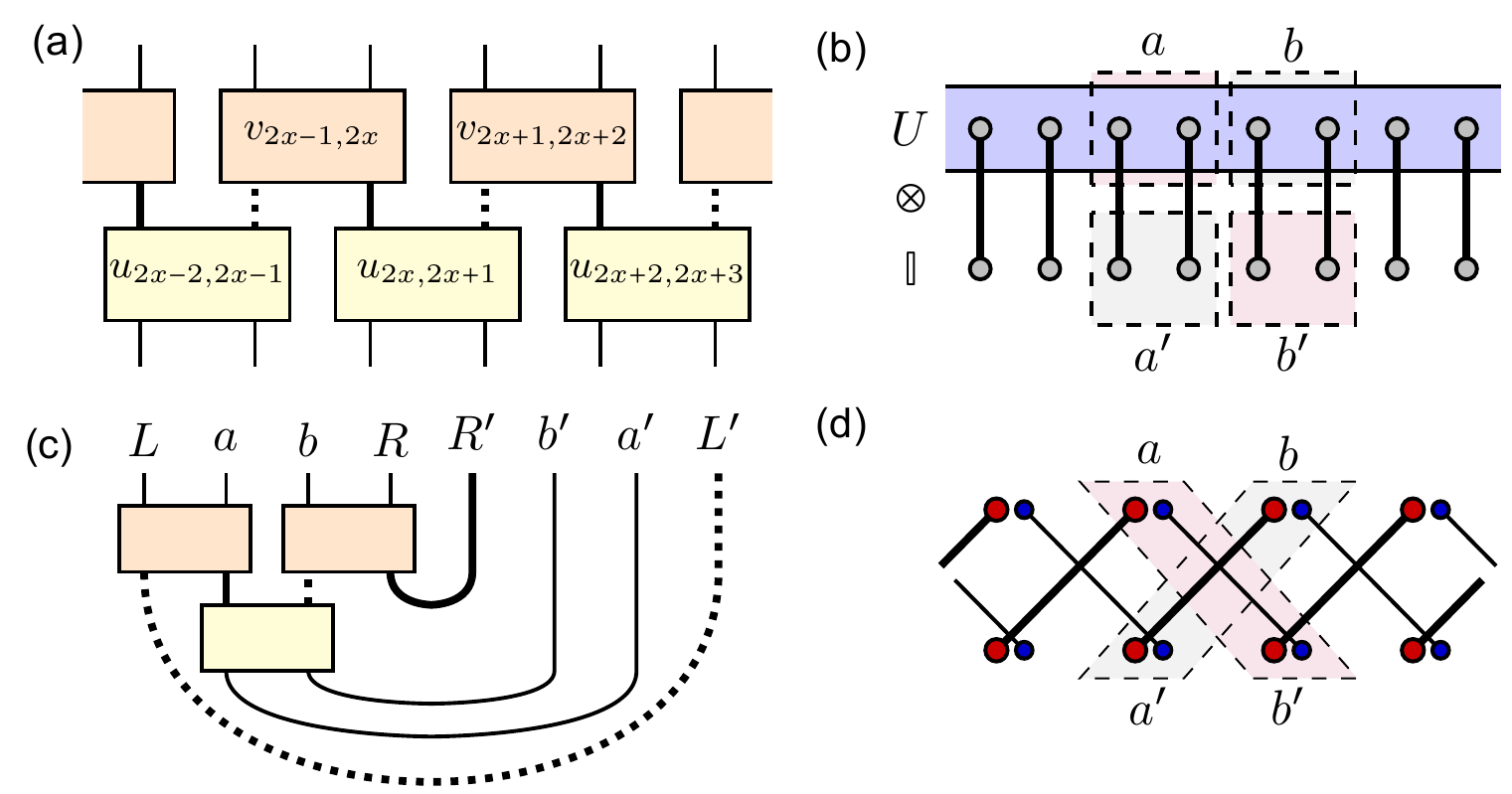}
       \end{center}
   \caption{(a) Representation of a QCA as a bilayer product of nearest neighbor unitaries, with possibly different Hilbert-space bipartition for their inputs and outputs. Here the thick and dotted legs refer to virtual local Hilbert spaces at even and odd sites in the hidden layer, respectively. (b) CJS $|U\rangle$ defined in Eq.~(\ref{CJS}) and the entanglement bipartition related to the index. (c) Relevant CJS used in the derivation of the entropy expression (\ref{dS}) and the main result (\ref{eq:main_result}). (d) Demostration of Eq.~(\ref{dS}) for a general representative QCA with index $\log(p/q)$ consisting of the right and left translations of local Hilbert spaces $\mathbb{C}^p$ (red circles) and $\mathbb{C}^q$ (blue circles), respectively.}
   \label{fig2}
\end{figure}

\emph{Index of a 1D QCA.---} We consider a general range-$r$ QCA $U$ defined on a periodic qudit chain with local Hilbert space $\mathbb{C}^d$ and length $N$. As shown in Refs.~\cite{schumacher2004reversible,cirac2017matrix}, by grouping (or ``blocking'') at least $r$ adjacent sites into one (such that the coarse-grained QCA has length $2l\le N/r$ and range-$1$), we can represent $U$ as [cf. Fig.~\ref{fig2}(a)] 
\begin{equation}
U=\left(\bigotimes^l_{x=1} v_{2x-1,2x}\right)\left(\bigotimes^l_{x=1} u_{2x,2x+1}\right)\,,
\end{equation}
where $u_{2x,2x+1}:\mathbb{C}^{d_{2x}}\otimes \mathbb{C}^{d_{2x+1}}\to \mathbb{C}^{d'_{2x}}\otimes \mathbb{C}^{d'_{2x+1}}$ and $v_{2x-1,2x}:\mathbb{C}^{d'_{2x-1}}\otimes \mathbb{C}^{d'_{2x}}\to \mathbb{C}^{d_{2x-1}}\otimes \mathbb{C}^{d_{2x}}$ are unitaries, with input and outp on two adjacent blocked sites, respectively. Note that the local Hilbert space dimension $d'_x$ in the ``hidden'' layer is not equal to $d_x$ in general, but they must satisfy $d'_{x}d'_{x+1}=d_xd_{x+1}$ $\forall x=1,2,...,2l$, implying $d'_{2x}/d_{2x}=d_{2x-1}/d'_{2x-1}$ is a constant independent of $x=1,2,...,l$. The index of $U$ is defined as the logarithm of this constant \cite{gross2012index,cirac2017matrix}:
\begin{equation}
\ind \equiv \log \frac{d_{2x}}{d'_{2x}}\in\log\mathbb{Q}^+,
\label{rank}
\end{equation}
which was proven to be stable against different ways of blocking and under continuous deformations. In other words, this index is a topological invariant of $U$.

\emph{An equivalent formulation of the  index.---} As a starting point of our work, we show that the index can be expressed exactly as an entanglement entropy difference between two reduced states of the vectorized operator $|U\rangle$, technically known as the Choi-Jamio{\l}kowski state (CJS)
\cite{Choi1975}:
\begin{equation}
|U\rangle\equiv (U\otimes \mathbb{I})|I\rangle,
\label{CJS}
\end{equation} 
where $|I\rangle \equiv d^{-N/2} (\sum^d_{j=1}|jj\rangle)^{\otimes N}$ is the maximally entangled state between two copies of the entire Hilbert space and $\mathbb{I}\equiv\openone^{\otimes N}$ is the global identity acting on the auxiliary copy. As shown in Fig.~\ref{fig2}(b), if we pick up two adjacent segments $a$ and $b$ with $\min\{|a|,|b|\}\ge r$ ($|a|$: number of sites in $a$) \footnote{To be rigorous, we should also require $|a|+|b|\le N-2r$, although typically $|a|,|b|\sim \mathcal{O}(r)\ll N$ is relevant to both practical numerical calculations and experimental measurements. The same constraint appears for the validity of Eq.~(\ref{eq:main_result}) and explains why the minimal experimental setup is $N=4$.}, irrespective of where they are located \cite{SM}, the index (\ref{rank}) turns out to be 
\begin{equation}
\ind =\frac{1}{2}(S_\alpha(\rho_{ab'})-S_\alpha(\rho_{a'b})),
\label{dS}
\end{equation}
where $\rho_{A}\equiv \Tr_{\bar A}[|U\rangle\langle U|]$ ($A=ab',a'b$ and $\bar{A}$ is the complement of $A$) and $S_\alpha(\rho)\equiv(1-\alpha)^{-1}\log\Tr[\rho^\alpha]$ can be an arbitrary R\'enyi entropy. Here and in the following, we denote by $a^\prime$ the ancillary qudits associated with $a$, and analogously for other regions. To show Eq.~(\ref{dS}), we can take a specific bilayer representation such that $a$ and $b$ are blocked into an even and odd site, respectively. Then we consider the CJS shown in Fig.~\ref{fig2}(c) corresponding to $(v_{L,a}\otimes v_{b,R}) u_{a,b}$, so that its reduced state on $Laa'bb'R$ coincides with that of $|U\rangle$. Importantly, here $L$ and $R$ are finite regions next to $a$, $b$ with $|L|, |R|\geq r$, cf. Fig.~\ref{fig2}(c). Since this is a pure state, we have
\begin{equation}
S_\alpha(\rho_{L'a'bR})=S_\alpha(\rho_{Lab'R'})
\label{gleich}
\end{equation}
under the bipartition $L'a'bR\sqcup La b'R'$. By directly contracting the tensor, we obtain
$\rho_{L'a'bR}=\frac{\openone_{L'}}{d'_L}\otimes \rho_{a'bR}$ and $\rho_{Lab'R'}=\rho_{Lab'}\otimes \frac{\openone_{R'}}{d'_R}$. Tracing out the auxiliary part of $|U\rangle$ except for $a'$, we can consider $\rho_{a'bR}$ as the reduced state of $(U\otimes\openone_{a'})(|I_{aa'}\rangle\langle I_{aa'}|\otimes \frac{d_a\mathbb{I}_{\bar a}}{d^N})(U^\dag\otimes\openone_{a'})$ ($|I_{aa'}\rangle$: maximally entangled state between $a$ and $a'$), which is supported on $Laa'b$. This implies $\rho_{a'bR}=\rho_{a'b}\otimes \frac{\openone_R}{d_R}$. Similarly, we can show $\rho_{Lab'}= \frac{\openone_L}{d_L}\otimes\rho_{ab'}$. Therefore, $\rho_{L'a'bR}$ and $\rho_{Lab'R'}$ turns out to be supported on $a'b$ and $ab'$, respectively:
\begin{equation}
\begin{split}
\rho_{L'a'bR}&=\frac{\openone_{L'}}{d'_L}\otimes \rho_{a'b}\otimes \frac{\openone_R}{d_R},\\
\rho_{Lab'R'}&=\frac{\openone_L}{d_L}\otimes \rho_{ab'}\otimes \frac{\openone_{R'}}{d'_R}.
\end{split}
\label{supp}
\end{equation}
Substituting Eq.~(\ref{supp}) into Eq.~(\ref{gleich}) and recalling the definition of index in Eq.~(\ref{rank}), we end up with Eq.~(\ref{dS}).

It is instructive to test Eq.~\eqref{dS} for the QCA shown in Fig.~\ref{fig2}(d), which are the simplest representatives with ${\rm ind}=\log(p/q)$. This example provides a nice illustration of the index as a measure of the chirality of quantum information flow.

Before proceeding, we compare our  result against different reformulations of the index which previously appeared in the literature. First, one can show~\cite{SM} that Eq.~\eqref{dS} is equivalent to the ``chiral mutual information'' introduced in Ref.~\cite{duschatko2018Tracking}, which is 
defined in terms of some local ancillary degrees of freedom. We stress, however, that the derivation presented there is completely different from the one reported here. In particular, we will see that our formalism naturally allows us to make a connection with different properties of the evolution operator. Second, the R\'enyi-2 version of Eq.~\eqref{dS} is easily seen to coincide with the original definition of the index, given in Ref.~\cite{gross2012index}, in terms algebra overlaps\cite{SM}.

\emph{Proof for the entanglement lower bound.---} Before proving Eq.~(\ref{eq:main_result}) for general R\'enyi-$\alpha$ entropies, we observe that one could show directly its validity for $\alpha= 1$, i.e. the case of von Neumann entropy, and  hence for $\alpha\leq 1$. Considering a segment with length larger than $2r$ as the subsystem $A$, we can bipartite it into two adjacent segments $A=a\sqcup b$ with $\min\{|a|,|b|\}\ge r$, so that the entropy formula (\ref{dS}) is valid. Since the von-Neumann entropy satisfies the triangle inequality \cite{Nielsen2010}, which follows from subadditivity, we have
 \begin{equation}
S(\rho_{aba'b'})\ge|S(\rho_{ab'})-S(\rho_{a'b})|=2|\ind |.
 \end{equation}
Using monotonicity in $\alpha$ of $S_{\alpha}$, we immediately get that the bound is satisfied for $\alpha\leq 1$. Unfortunately, this proof cannot be extended to $\alpha > 1$,  since the R\'enyi entropies do not satisfy subadditivity in general \cite{Linden2013}. 

To prove the general case, we should make further use of some nice properties of $\rho_{AA'}$ as a reduced state of a pure CJS. To this end, let us return to the state shown in Fig.~\ref{fig2}(c) and take a different bipartition $aba'b'\sqcup LRL'R'$, obtaining
\begin{equation}
S_\alpha(\rho_{aba'b'})=S_\alpha(\rho_{LRL'R'})=S_\alpha(\rho_{LL'}) +S_\alpha(\rho_{RR'}),
\end{equation}
where we have used $\rho_{LRL'R'}=\rho_{LL'}\otimes\rho_{RR'}$. This relation follows from the fact that $U$ is a QCA~\cite{piroli2020quantum}, and can be understood from the vanishing correlation for two arbitrary observables supported on these two regions~\cite{SM}. While the R\'enyi entropies do not satisfy subadditivity, they still satisfy the weak subadditivity \cite{Hayden2002}: 
\begin{equation}
S_\alpha(\rho_{LL'})\ge \max\{S_\alpha(\rho_L) -S_0(\rho_{L'}),S_\alpha(\rho_{L'}) -S_0(\rho_L)\},
\end{equation}
where $S_0(\rho)\equiv\log({\rm rank}\rho)$ is also called the Max entropy in the sense that it is the largest in the R\'enyi family. This inequality can be shown from the non-decreasing property of R\'enyi entropies upon arbitrary unital channels \cite{Hayden2002}. Note that $\rho_L$ and $\rho_{L'}$ are both maximally mixed and thus their R\'enyi entropies coincide with the maximum possible values $\log d_L$ and $\log d'_L$, respectively. Therefore, we obtain $S_\alpha(\rho_{LL'})\ge |\ind|$. Similarly, we have $S_\alpha(\rho_{RR'})\ge |\ind|$ and thus $S_\alpha(\rho_{aba'b'})\ge 2|\ind |$, which completes the proof of the main result.

Obviously, the bound is tight for all the R\'enyi entropies for $|\ind| \in \log\mathbb{Z}^+$ since they are saturated by left or right translations. What is less clear is whether the bound is tight for general $\ind \in\log \mathbb{Q}^+$. At least for the R\'enyi-$\infty$ entropy, which gives the strongest version of Eq.~(\ref{eq:main_result}) for a given QCA, we can readily construct an example which  saturates the bound. To this end, we take $u:\mathbb{C}^p\otimes\mathbb{C}^q\to\mathbb{C}^q\otimes \mathbb{C}^p$ (with $q>p$) in Fig.~\ref{fig1}(c), to be
\begin{equation}
u=\sum^p_{m,n=1}|mn\rangle\langle mn|+\sum^p_{m=1}\sum^q_{n=p+1}|nm\rangle\langle mn|.
\end{equation}
This construction is reminiscent of the AKLT state \cite{Tasaki1987}, in the sense that it is essentially an assembly of disjoint unitaries but becomes correlated upon recombinations of subsystems.
See Fig.~\ref{fig1}(d) for a demonstration for $(p,q)=(2,3)$. On the other hand, since $S_\alpha >S_\infty$ for any finite $\alpha$ and $S_\infty\in \log(\mathbb{Q}\backslash\mathbb{Z}^+)$, the bound is not tight for any noninteger index and $\alpha<\infty$. Identifying a tight bound for the most general case thus remains an open problem.

One immediate and important implication of our main result is that it rigorously rules out, for QCAs with nonzero index, the possibility of many-body localization, which implies a logarithmic growth of the entanglement of the evolution operator~\cite{Zhou2017}. More generally, for nonzero index, any diffusive behavior characterized by a sublinear growth of the R\'enyi entropies, as recently demonstrated for quantum-state entanglement in random circuits with a diffusive charge~\cite{huang2019dynamics,rakovszky2019subballistic}, is forbidden. This is because the index is additive upon compositions \cite{gross2012index} so the operator entanglement entropy after $t$ time steps is lower bounded by \footnote{To avoid undesired finite-size saturation, we may choose the subsystem size to be no smaller than $2rt$ for a range-$r$ QCA at time $t$; alternatively, given subsystem size $|A|$, we should focus on a time interval up to $|A|/(2r)$.}
\begin{equation}\label{eq:bound_time}
S_\alpha(t)\ge 2|\ind |t,
\end{equation} 
implying a linear growth. When specified to the case $\alpha=2$, this result also allows to make a precise connection to quantum chaos, due to the known relation between $S_2(t)$ and the average of infinite-temperature OTOCs \cite{hosur2016chaos}: 
\begin{equation}\label{eq:otocs}
\overline{\langle U^tO_A(U^{\dag })^t O_{\bar A}U^tO_A(U^{\dag} )^t O_{\bar A}\rangle}_{\beta=0} = e^{-S_2(t)}.
\end{equation}
Here the l.h.s. denotes the normalized sum over the elements $O_A$, $O_{\bar{A}}$ of a complete basis of observables supported in $A$, $\bar{A}$, respectively~\cite{hosur2016chaos}. When combined with  Eq.~\eqref{eq:bound_time}, we immediately obtain that a nonzero index implies exponential decay of the averaged OTOC. Once again, this is not consistent with an MBL evolution, for which OTOCs exhibit a power-law decay \cite{Fan2017,Chen2017}. Finally, we remark that there is no universal lower bound for the growth of the entanglement of an initial product state evolved by a QCA, as it can be seen in the simple case of translations.

\emph{Experimental relevance.---} One important motivation to consider R\'enyi entropies is their accessibility in state-of-the-art quantum simulation experiments \cite{Greiner2015,Greiner2016,Linke2018,Greiner2019}. Thanks to the reformulation~\eqref{dS} of the index in terms of entropies, all the quantities in our main result (\ref{eq:main_result}) can be in principle measured.

\begin{figure}
\begin{center}
       \includegraphics[width=8.5cm, clip]{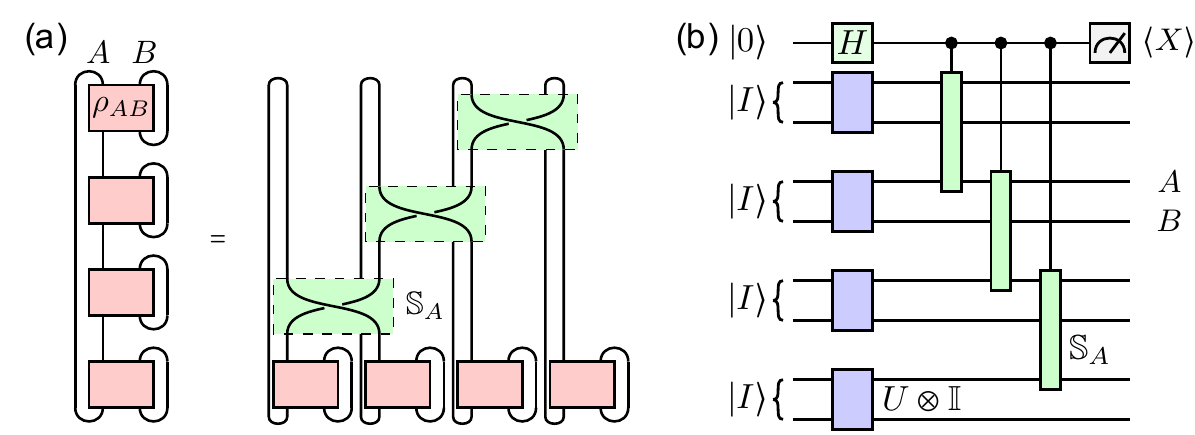}
       \end{center}
   \caption{(a) Graphic representation of $\Tr[\rho_A^n]=\Tr[\mathbb{T}_A\rho_{AB}^{\otimes n}]$ with $\mathbb{T}_A=\mathbb{S}^{[n-1,n]}_A...\mathbb{S}^{[2,3]}_A\mathbb{S}^{[1,2]}_A$, where $\mathbb{S}^{[j,k]}_A$ is a SWAP over the $j$th and $k$th copies of subsystem $A$.  (b) Experimental setup for measuring the operator entanglement R\'enyi-$n$ entropy as well as the index of a QCA. Here $\mathbb{S}_A$'s are controlled by the ancilla qubit, $H=(X+Z)/\sqrt{2}$ is the Hadamard gate, $|I\rangle$'s are global maximally entangled states which become $|U\rangle$ upon the action of $U$, and $A$ is a subsystem of interest, whose choice depends on which quantity (index or operator entanglement) we would like to measure. }
   \label{fig3}
\end{figure}

Let us explain the protocol to measure R\'enyi-$n$ entropies with $n\in\mathbb{N}$. We can straightforwardly generalize the strategies in Refs.~\cite{Ekert2002,Abanin2012,Daley2012} for quantum states to operators by vectorizing the latter into their CJSs. To measure $S_n$ of a bipartite state $\rho_{AB}=|\Psi_{AB}\rangle\langle\Psi_{AB}|$, the essential idea is make use of $\Tr[\rho_A^n]=\Tr[\mathbb{T}_A\rho_{AB}^{\otimes n}]$, where $\mathbb{T}_A$ denotes the translation operator of subsystems $A$, which can be implemented by a sequence of SWAP unitaries [cf. Fig.~\ref{fig3}(a)]. The value of $S_{n}$ can then be extracted by performing an interferometric measurement on an ancilla qubit after a sequence of controlled-SWAP gates as shown in Fig.~\ref{fig3}(b)~\cite{Ekert2002}. For our specific purpose of measuring the operator entanglement and the index, we should choose $A$ to be $aa'bb'$, $ab'$ and $a'b$, where $a$ and $b$ are adjacent segments whose lengths can be as small as the range of the QCA. In a proof-of-principle experiment, it would be good enough to construct a translation of qubits with $N=4$ and $r=1$, which may be implemented by $3$ SWAP gates, so that we can set $|a|=1$ and $|b|=1$, and measure the R\'enyi-2 entropy so that we only need a single controlled-SWAP gate acting on $2$ (index) or $4$ (operator entanglement) pairs of qubits. This minimal setup should be accessible by many current experimental platforms  such as trapped ions \cite{Blatt2012}, superconducting qubits \cite{Koch2012} and Rydberg-atom arrays \cite{Levine2018}. Note that there is also a more sophisticated method of measuring R\'enyi-$n$ entropies based on random quenches \cite{Elben2018,Vermersch2018} which has also been experimentally realized \cite{Roos2019} and should also be applicable to measuring operator entanglement and the index. As a final remark, we mention that one can further reduce the experimental cost in state preparation by taking advantage of the idea in Ref.~\cite{duschatko2018Tracking}. That is, we may only introduce a few ancillas covering the subsystem, while obtaining the same measurement results.

\emph{Stability against exponential tails.---} 
Trivial (nontrivial) QCAs have been used to approximate the (edge) dynamics of 1D Floquet unitaries \cite{Osborne2006} (2D chiral Floquet MBL phases \cite{Po2016}) governed by time-periodic local Hamiltonians satisfying the Lieb-Robinson bound \cite{Lieb1972,Bravyi2006,Nachtergaele2006}. The QCA approximation should thus be precise up to some exponential tails outside the light cone. This motivates us to analyze how our main result (\ref{eq:main_result}) is modified by such small deviation from QCAs.

To avoid the problem of defining the index for quasi-local unitaries with exponential tails, which remains an open problem \footnote{Very recently, this is partially solved in Ref.~\cite{Ranard2020} for infinite and finite open chains, but remains unsolved for finite periodic systems discussed here.}, we restrict ourselves to consider specific quasi-local unitaries that are range-$r$ QCAs followed by finite-time evolutions of local Hamiltoniains, i.e., 
\begin{equation}
U= \hat{\rm T}e^{-i\int^T_0dt H(t)}U_{\rm QCA},
\label{UH}
\end{equation} 
where $\hat{\rm T}$ denotes the time ordering, $H(t)=\sum^N_{j=1} h_j(t)$ with $h_j(t)$ supported locally near $j$ and $h\equiv\max_{j,t}\|h_j(t)\|$ is finite. This setup has been used in several previous studies \cite{Po2016,duschatko2018Tracking}. Setting the rhs of Eq.~(\ref{eq:main_result}) as the index of $U_{\rm QCA}$, we would like to know whether the inequality can be violated and, if yes, to what extent. 

We can give an explicit example where the inequality is violated by choosing $U_{\rm QCA}$ to be the right translation $\mathbb{T}$ and $H(t)=h \mathbb{S}^{[j_t,j_t+1]}$, where $j_t= \lfloor t |A|/T\rfloor$ and $A=[0,|A|-1]$ is the subsystem of interest, while $\mathbb{S}^{[j,k]}$ is the SWAP operator between sites $j$ and $k$. This exactly solvable construction is inspired by the fact that $AA'$ and $\bar A\bar A'$ can be exactly disentangled for $T=\pi |A|/(2h)$, so one may expect the Hamiltonian evolution is still a disentangler for a finite $T\ll |A|$. Indeed, we find that Eq.~(\ref{eq:main_result}) is violated for all $\alpha>0$, with the largest violation being $2\ind-S_\infty =\log[1+(d^2-1)\epsilon]>0$,  where $\epsilon =\sin^{2|A|}(hT/|A|)$ scales as $e^{-\mathcal{O}(|A|\log |A|)}$ for large $|A|$, which also determines the scaling behavior of the violation.

In fact, the above example serves as a qualitatively worst case. That is, we can prove that for any local-Hamiltonian evolution, the order of the violation of Eq.~(\ref{eq:main_result}) can never be larger than $e^{-\mathcal{O}(|A|\log |A|)}$ \cite{SM}, and thus vanishes superexponentially in the thermodynamic limit. The proof involves a technique in Ref.~\cite{Osborne2006} for approximating Hamiltonian dynamics by quantum circuits and a careful optimization of the Lieb-Robinson bound \cite{Hastings2010,Gong2020}. This rigorous derivation implies the stability of our main result and significantly widens its range of applicability. For example, we can rule out any sublinear entanglement-growth behavior for the evolution operator $e^{-iHT}\mathbb{T}$ (thus forbidding MBL features), even if the local Hamiltonian $H$ is in the deep MBL phase \cite{Po2016}. 

\emph{Summary and outlook.---} We have derived a convenient local expression for the index of a 1D QCA, and proved that any R\'enyi-$\alpha$ entropy of the evolution operator is lower-bounded by twice the index. This rigorous bound rules out any sublinear entanglement-growth behavoir in nontrivial QCAs and might be interpreted as a lower bound on quantum chaos, as opposed to the Maldacena-Shenker-Stanford upper bound \cite{Stanford2016}. Since the R\'enyi entropy is accessible in cutting edge AMO experiments, our results should be experimentally observable. We have also discussed the validity of our bound against deviations from QCAs by exponential tails.

One immediate question for future work is how to tighten the bound for a general rational index and R\'enyi-$\alpha$ entropy with $\alpha<\infty$. Another natural direction to explore is the generalization to the symmetry-protected case~\cite{gong2020classification}. Here we expect that a nonzero symmetry-protected index will give rise to a linear growth of the entropy, even for $\ind=0$. Finally, it may also be interesting to consider generalizations to quantum channels with suitable locality properties \cite{piroli2020quantum}.

\emph{Note added.}--- While finalizing this manuscript, a related work by Ranard \emph{et al.} appeared in Ref.~\cite{Ranard2020}, which also reported the entropy reformulation of the index. However, their main focus was to derive an index theory for quasi-local unitaries rather than exploring its connections with other dynamical properties. On the technical level, Ref.~\cite{Ranard2020} only uses the von Neumann entropy for infinite (or finite open) chains. 

We thank Andrew C. Potter for very helpful comments on the manuscript. Z.G. is supported by the Max-Planck-Harvard Research Center for Quantum Optics (MPHQ). L. P. acknowledges support from the Alexander von Humboldt foundation. J. I. C. acknowledges support by the EU Horizon 2020 program through the ERC Advanced Grant QENOCOBA No. 742102 and from the DFG (German Research Foundation) under Germany’s Excellence Strategy--- EXC-2111---390814868.

\bibliography{GZP_references}

\begin{thebibliography}{76}%
\makeatletter
\providecommand \@ifxundefined [1]{%
 \@ifx{#1\undefined}
}%
\providecommand \@ifnum [1]{%
 \ifnum #1\expandafter \@firstoftwo
 \else \expandafter \@secondoftwo
 \fi
}%
\providecommand \@ifx [1]{%
 \ifx #1\expandafter \@firstoftwo
 \else \expandafter \@secondoftwo
 \fi
}%
\providecommand \natexlab [1]{#1}%
\providecommand \enquote  [1]{``#1''}%
\providecommand \bibnamefont  [1]{#1}%
\providecommand \bibfnamefont [1]{#1}%
\providecommand \citenamefont [1]{#1}%
\providecommand \href@noop [0]{\@secondoftwo}%
\providecommand \href [0]{\begingroup \@sanitize@url \@href}%
\providecommand \@href[1]{\@@startlink{#1}\@@href}%
\providecommand \@@href[1]{\endgroup#1\@@endlink}%
\providecommand \@sanitize@url [0]{\catcode `\\12\catcode `\$12\catcode
  `\&12\catcode `\#12\catcode `\^12\catcode `\_12\catcode `\%12\relax}%
\providecommand \@@startlink[1]{}%
\providecommand \@@endlink[0]{}%
\providecommand \url  [0]{\begingroup\@sanitize@url \@url }%
\providecommand \@url [1]{\endgroup\@href {#1}{\urlprefix }}%
\providecommand \urlprefix  [0]{URL }%
\providecommand \Eprint [0]{\href }%
\providecommand \doibase [0]{http://dx.doi.org/}%
\providecommand \selectlanguage [0]{\@gobble}%
\providecommand \bibinfo  [0]{\@secondoftwo}%
\providecommand \bibfield  [0]{\@secondoftwo}%
\providecommand \translation [1]{[#1]}%
\providecommand \BibitemOpen [0]{}%
\providecommand \bibitemStop [0]{}%
\providecommand \bibitemNoStop [0]{.\EOS\space}%
\providecommand \EOS [0]{\spacefactor3000\relax}%
\providecommand \BibitemShut  [1]{\csname bibitem#1\endcsname}%
\let\auto@bib@innerbib\@empty
\bibitem [{\citenamefont {Schumacher}\ and\ \citenamefont
  {Werner}(2004)}]{schumacher2004reversible}%
  \BibitemOpen
  \bibfield  {author} {\bibinfo {author} {\bibfnamefont {B.}~\bibnamefont
  {Schumacher}}\ and\ \bibinfo {author} {\bibfnamefont {R.~F.}\ \bibnamefont
  {Werner}},\ }\href@noop {} {\bibfield  {journal} {\bibinfo  {journal} {arXiv
  quant-ph/0405174}\ } (\bibinfo {year} {2004})}\BibitemShut {NoStop}%
\bibitem [{\citenamefont {Farrelly}(2019)}]{farrelly2019review}%
  \BibitemOpen
  \bibfield  {author} {\bibinfo {author} {\bibfnamefont {T.}~\bibnamefont
  {Farrelly}},\ }\href@noop {} {\bibfield  {journal} {\bibinfo  {journal}
  {arXiv:1904.13318}\ } (\bibinfo {year} {2019})}\BibitemShut {NoStop}%
\bibitem [{\citenamefont {Arrighi}(2019)}]{arrighi2019overview}%
  \BibitemOpen
  \bibfield  {author} {\bibinfo {author} {\bibfnamefont {P.}~\bibnamefont
  {Arrighi}},\ }\href {\doibase 10.1007/s11047-019-09762-6} {\bibfield
  {journal} {\bibinfo  {journal} {Natural Comp.}\ }\textbf {\bibinfo {volume}
  {18}},\ \bibinfo {pages} {885} (\bibinfo {year} {2019})}\BibitemShut
  {NoStop}%
\bibitem [{\citenamefont {Nahum}\ \emph {et~al.}(2017)\citenamefont {Nahum},
  \citenamefont {Ruhman}, \citenamefont {Vijay},\ and\ \citenamefont
  {Haah}}]{nahum2017quantum}%
  \BibitemOpen
  \bibfield  {author} {\bibinfo {author} {\bibfnamefont {A.}~\bibnamefont
  {Nahum}}, \bibinfo {author} {\bibfnamefont {J.}~\bibnamefont {Ruhman}},
  \bibinfo {author} {\bibfnamefont {S.}~\bibnamefont {Vijay}}, \ and\ \bibinfo
  {author} {\bibfnamefont {J.}~\bibnamefont {Haah}},\ }\href {\doibase
  10.1103/PhysRevX.7.031016} {\bibfield  {journal} {\bibinfo  {journal} {Phys.
  Rev. X}\ }\textbf {\bibinfo {volume} {7}},\ \bibinfo {pages} {031016}
  (\bibinfo {year} {2017})}\BibitemShut {NoStop}%
\bibitem [{\citenamefont {Nahum}\ \emph {et~al.}(2018)\citenamefont {Nahum},
  \citenamefont {Vijay},\ and\ \citenamefont {Haah}}]{nahum2018operator}%
  \BibitemOpen
  \bibfield  {author} {\bibinfo {author} {\bibfnamefont {A.}~\bibnamefont
  {Nahum}}, \bibinfo {author} {\bibfnamefont {S.}~\bibnamefont {Vijay}}, \ and\
  \bibinfo {author} {\bibfnamefont {J.}~\bibnamefont {Haah}},\ }\href {\doibase
  10.1103/PhysRevX.8.021014} {\bibfield  {journal} {\bibinfo  {journal} {Phys.
  Rev. X}\ }\textbf {\bibinfo {volume} {8}},\ \bibinfo {pages} {021014}
  (\bibinfo {year} {2018})}\BibitemShut {NoStop}%
\bibitem [{\citenamefont {von Keyserlingk}\ \emph {et~al.}(2018)\citenamefont
  {von Keyserlingk}, \citenamefont {Rakovszky}, \citenamefont {Pollmann},\ and\
  \citenamefont {Sondhi}}]{vonKeyserlingk2018operator}%
  \BibitemOpen
  \bibfield  {author} {\bibinfo {author} {\bibfnamefont {C.~W.}\ \bibnamefont
  {von Keyserlingk}}, \bibinfo {author} {\bibfnamefont {T.}~\bibnamefont
  {Rakovszky}}, \bibinfo {author} {\bibfnamefont {F.}~\bibnamefont {Pollmann}},
  \ and\ \bibinfo {author} {\bibfnamefont {S.~L.}\ \bibnamefont {Sondhi}},\
  }\href {\doibase 10.1103/PhysRevX.8.021013} {\bibfield  {journal} {\bibinfo
  {journal} {Phys. Rev. X}\ }\textbf {\bibinfo {volume} {8}},\ \bibinfo {pages}
  {021013} (\bibinfo {year} {2018})}\BibitemShut {NoStop}%
\bibitem [{\citenamefont {Rakovszky}\ \emph {et~al.}(2018)\citenamefont
  {Rakovszky}, \citenamefont {Pollmann},\ and\ \citenamefont {von
  Keyserlingk}}]{Tibor2018}%
  \BibitemOpen
  \bibfield  {author} {\bibinfo {author} {\bibfnamefont {T.}~\bibnamefont
  {Rakovszky}}, \bibinfo {author} {\bibfnamefont {F.}~\bibnamefont {Pollmann}},
  \ and\ \bibinfo {author} {\bibfnamefont {C.~W.}\ \bibnamefont {von
  Keyserlingk}},\ }\href {\doibase 10.1103/PhysRevX.8.031058} {\bibfield
  {journal} {\bibinfo  {journal} {Phys. Rev. X}\ }\textbf {\bibinfo {volume}
  {8}},\ \bibinfo {pages} {031058} (\bibinfo {year} {2018})}\BibitemShut
  {NoStop}%
\bibitem [{\citenamefont {Khemani}\ \emph {et~al.}(2018)\citenamefont
  {Khemani}, \citenamefont {Vishwanath},\ and\ \citenamefont
  {Huse}}]{Khemani2018}%
  \BibitemOpen
  \bibfield  {author} {\bibinfo {author} {\bibfnamefont {V.}~\bibnamefont
  {Khemani}}, \bibinfo {author} {\bibfnamefont {A.}~\bibnamefont {Vishwanath}},
  \ and\ \bibinfo {author} {\bibfnamefont {D.~A.}\ \bibnamefont {Huse}},\
  }\href {\doibase 10.1103/PhysRevX.8.031057} {\bibfield  {journal} {\bibinfo
  {journal} {Phys. Rev. X}\ }\textbf {\bibinfo {volume} {8}},\ \bibinfo {pages}
  {031057} (\bibinfo {year} {2018})}\BibitemShut {NoStop}%
\bibitem [{\citenamefont {S\"underhauf}\ \emph {et~al.}(2018)\citenamefont
  {S\"underhauf}, \citenamefont {P\'erez-Garc\'{\i}a}, \citenamefont {Huse},
  \citenamefont {Schuch},\ and\ \citenamefont {Cirac}}]{Christoph2018}%
  \BibitemOpen
  \bibfield  {author} {\bibinfo {author} {\bibfnamefont {C.}~\bibnamefont
  {S\"underhauf}}, \bibinfo {author} {\bibfnamefont {D.}~\bibnamefont
  {P\'erez-Garc\'{\i}a}}, \bibinfo {author} {\bibfnamefont {D.~A.}\
  \bibnamefont {Huse}}, \bibinfo {author} {\bibfnamefont {N.}~\bibnamefont
  {Schuch}}, \ and\ \bibinfo {author} {\bibfnamefont {J.~I.}\ \bibnamefont
  {Cirac}},\ }\href {\doibase 10.1103/PhysRevB.98.134204} {\bibfield  {journal}
  {\bibinfo  {journal} {Phys. Rev. B}\ }\textbf {\bibinfo {volume} {98}},\
  \bibinfo {pages} {134204} (\bibinfo {year} {2018})}\BibitemShut {NoStop}%
\bibitem [{\citenamefont {Chan}\ \emph
  {et~al.}(2018{\natexlab{a}})\citenamefont {Chan}, \citenamefont {De~Luca},\
  and\ \citenamefont {Chalker}}]{Chan2018}%
  \BibitemOpen
  \bibfield  {author} {\bibinfo {author} {\bibfnamefont {A.}~\bibnamefont
  {Chan}}, \bibinfo {author} {\bibfnamefont {A.}~\bibnamefont {De~Luca}}, \
  and\ \bibinfo {author} {\bibfnamefont {J.~T.}\ \bibnamefont {Chalker}},\
  }\href {\doibase 10.1103/PhysRevLett.121.060601} {\bibfield  {journal}
  {\bibinfo  {journal} {Phys. Rev. Lett.}\ }\textbf {\bibinfo {volume} {121}},\
  \bibinfo {pages} {060601} (\bibinfo {year} {2018}{\natexlab{a}})}\BibitemShut
  {NoStop}%
\bibitem [{\citenamefont {Chan}\ \emph
  {et~al.}(2018{\natexlab{b}})\citenamefont {Chan}, \citenamefont {De~Luca},\
  and\ \citenamefont {Chalker}}]{Chan2018b}%
  \BibitemOpen
  \bibfield  {author} {\bibinfo {author} {\bibfnamefont {A.}~\bibnamefont
  {Chan}}, \bibinfo {author} {\bibfnamefont {A.}~\bibnamefont {De~Luca}}, \
  and\ \bibinfo {author} {\bibfnamefont {J.~T.}\ \bibnamefont {Chalker}},\
  }\href {\doibase 10.1103/PhysRevX.8.041019} {\bibfield  {journal} {\bibinfo
  {journal} {Phys. Rev. X}\ }\textbf {\bibinfo {volume} {8}},\ \bibinfo {pages}
  {041019} (\bibinfo {year} {2018}{\natexlab{b}})}\BibitemShut {NoStop}%
\bibitem [{\citenamefont {Friedman}\ \emph {et~al.}(2019)\citenamefont
  {Friedman}, \citenamefont {Chan}, \citenamefont {De~Luca},\ and\
  \citenamefont {Chalker}}]{Friedman2019}%
  \BibitemOpen
  \bibfield  {author} {\bibinfo {author} {\bibfnamefont {A.~J.}\ \bibnamefont
  {Friedman}}, \bibinfo {author} {\bibfnamefont {A.}~\bibnamefont {Chan}},
  \bibinfo {author} {\bibfnamefont {A.}~\bibnamefont {De~Luca}}, \ and\
  \bibinfo {author} {\bibfnamefont {J.~T.}\ \bibnamefont {Chalker}},\ }\href
  {\doibase 10.1103/PhysRevLett.123.210603} {\bibfield  {journal} {\bibinfo
  {journal} {Phys. Rev. Lett.}\ }\textbf {\bibinfo {volume} {123}},\ \bibinfo
  {pages} {210603} (\bibinfo {year} {2019})}\BibitemShut {NoStop}%
\bibitem [{\citenamefont {Bertini}\ \emph {et~al.}(2019)\citenamefont
  {Bertini}, \citenamefont {Kos},\ and\ \citenamefont {Prosen}}]{Bruno2019}%
  \BibitemOpen
  \bibfield  {author} {\bibinfo {author} {\bibfnamefont {B.}~\bibnamefont
  {Bertini}}, \bibinfo {author} {\bibfnamefont {P.}~\bibnamefont {Kos}}, \ and\
  \bibinfo {author} {\bibfnamefont {T.}~\bibnamefont {Prosen}},\ }\href
  {\doibase 10.1103/PhysRevLett.123.210601} {\bibfield  {journal} {\bibinfo
  {journal} {Phys. Rev. Lett.}\ }\textbf {\bibinfo {volume} {123}},\ \bibinfo
  {pages} {210601} (\bibinfo {year} {2019})}\BibitemShut {NoStop}%
\bibitem [{\citenamefont {Bertini}\ and\ \citenamefont
  {Piroli}(2020)}]{Bruno2020}%
  \BibitemOpen
  \bibfield  {author} {\bibinfo {author} {\bibfnamefont {B.}~\bibnamefont
  {Bertini}}\ and\ \bibinfo {author} {\bibfnamefont {L.}~\bibnamefont
  {Piroli}},\ }\href {\doibase 10.1103/PhysRevB.102.064305} {\bibfield
  {journal} {\bibinfo  {journal} {Phys. Rev. B}\ }\textbf {\bibinfo {volume}
  {102}},\ \bibinfo {pages} {064305} (\bibinfo {year} {2020})}\BibitemShut
  {NoStop}%
\bibitem [{\citenamefont {Xu}\ and\ \citenamefont
  {Swingle}(2020)}]{Swingle2020}%
  \BibitemOpen
  \bibfield  {author} {\bibinfo {author} {\bibfnamefont {S.}~\bibnamefont
  {Xu}}\ and\ \bibinfo {author} {\bibfnamefont {B.}~\bibnamefont {Swingle}},\
  }\href {https://doi.org/10.1038/s41567-019-0712-4} {\bibfield  {journal}
  {\bibinfo  {journal} {Nat. Phys.}\ }\textbf {\bibinfo {volume} {16}},\
  \bibinfo {pages} {199 } (\bibinfo {year} {2020})}\BibitemShut {NoStop}%
\bibitem [{\citenamefont {Claeys}\ and\ \citenamefont
  {Lamacraft}(2020)}]{Claeys2020Maximum}%
  \BibitemOpen
  \bibfield  {author} {\bibinfo {author} {\bibfnamefont {P.~W.}\ \bibnamefont
  {Claeys}}\ and\ \bibinfo {author} {\bibfnamefont {A.}~\bibnamefont
  {Lamacraft}},\ }\href {\doibase 10.1103/PhysRevResearch.2.033032} {\bibfield
  {journal} {\bibinfo  {journal} {Phys. Rev. Research}\ }\textbf {\bibinfo
  {volume} {2}},\ \bibinfo {pages} {033032} (\bibinfo {year}
  {2020})}\BibitemShut {NoStop}%
\bibitem [{\citenamefont {Arrighi}\ \emph {et~al.}(2011)\citenamefont
  {Arrighi}, \citenamefont {Nesme},\ and\ \citenamefont
  {Werner}}]{arrighi2011unitarity}%
  \BibitemOpen
  \bibfield  {author} {\bibinfo {author} {\bibfnamefont {P.}~\bibnamefont
  {Arrighi}}, \bibinfo {author} {\bibfnamefont {V.}~\bibnamefont {Nesme}}, \
  and\ \bibinfo {author} {\bibfnamefont {R.}~\bibnamefont {Werner}},\ }\href
  {\doibase 10.1016/j.jcss.2010.05.004} {\bibfield  {journal} {\bibinfo
  {journal} {J. Comp. Syst. Sciences}\ }\textbf {\bibinfo {volume} {77}},\
  \bibinfo {pages} {372} (\bibinfo {year} {2011})}\BibitemShut {NoStop}%
\bibitem [{\citenamefont {Haah}\ \emph {et~al.}(2018)\citenamefont {Haah},
  \citenamefont {Fidkowski},\ and\ \citenamefont
  {Hastings}}]{haah2018nontrivial}%
  \BibitemOpen
  \bibfield  {author} {\bibinfo {author} {\bibfnamefont {J.}~\bibnamefont
  {Haah}}, \bibinfo {author} {\bibfnamefont {L.}~\bibnamefont {Fidkowski}}, \
  and\ \bibinfo {author} {\bibfnamefont {M.~B.}\ \bibnamefont {Hastings}},\
  }\href@noop {} {\bibfield  {journal} {\bibinfo  {journal} {arXiv:1812.01625}\
  } (\bibinfo {year} {2018})}\BibitemShut {NoStop}%
\bibitem [{\citenamefont {Haah}(2019)}]{haah2019clifford}%
  \BibitemOpen
  \bibfield  {author} {\bibinfo {author} {\bibfnamefont {J.}~\bibnamefont
  {Haah}},\ }\href@noop {} {\bibfield  {journal} {\bibinfo  {journal}
  {arXiv:1907.02075}\ } (\bibinfo {year} {2019})}\BibitemShut {NoStop}%
\bibitem [{\citenamefont {Freedman}\ and\ \citenamefont
  {Hastings}(2020)}]{freedman2020classification}%
  \BibitemOpen
  \bibfield  {author} {\bibinfo {author} {\bibfnamefont {M.}~\bibnamefont
  {Freedman}}\ and\ \bibinfo {author} {\bibfnamefont {M.~B.}\ \bibnamefont
  {Hastings}},\ }\href {\doibase 10.1007/s00220-020-03735-y} {\bibfield
  {journal} {\bibinfo  {journal} {Comm. Math. Phys.}\ }\textbf {\bibinfo
  {volume} {376}},\ \bibinfo {pages} {1171} (\bibinfo {year}
  {2020})}\BibitemShut {NoStop}%
\bibitem [{\citenamefont {Piroli}\ and\ \citenamefont
  {Cirac}(2020)}]{piroli2020quantum}%
  \BibitemOpen
  \bibfield  {author} {\bibinfo {author} {\bibfnamefont {L.}~\bibnamefont
  {Piroli}}\ and\ \bibinfo {author} {\bibfnamefont {J.~I.}\ \bibnamefont
  {Cirac}},\ }\href {\doibase 10.1103/PhysRevLett.125.190402} {\bibfield
  {journal} {\bibinfo  {journal} {Phys. Rev. Lett.}\ }\textbf {\bibinfo
  {volume} {125}},\ \bibinfo {pages} {190402} (\bibinfo {year}
  {2020})}\BibitemShut {NoStop}%
\bibitem [{\citenamefont {Gross}\ \emph {et~al.}(2012)\citenamefont {Gross},
  \citenamefont {Nesme}, \citenamefont {Vogts},\ and\ \citenamefont
  {Werner}}]{gross2012index}%
  \BibitemOpen
  \bibfield  {author} {\bibinfo {author} {\bibfnamefont {D.}~\bibnamefont
  {Gross}}, \bibinfo {author} {\bibfnamefont {V.}~\bibnamefont {Nesme}},
  \bibinfo {author} {\bibfnamefont {H.}~\bibnamefont {Vogts}}, \ and\ \bibinfo
  {author} {\bibfnamefont {R.~F.}\ \bibnamefont {Werner}},\ }\href {\doibase
  10.1007/s00220-012-1423-1} {\bibfield  {journal} {\bibinfo  {journal} {Comm.
  Math. Phys.}\ }\textbf {\bibinfo {volume} {310}},\ \bibinfo {pages} {419}
  (\bibinfo {year} {2012})}\BibitemShut {NoStop}%
\bibitem [{\citenamefont {Cirac}\ \emph {et~al.}(2017)\citenamefont {Cirac},
  \citenamefont {Perez-Garcia}, \citenamefont {Schuch},\ and\ \citenamefont
  {Verstraete}}]{cirac2017matrix}%
  \BibitemOpen
  \bibfield  {author} {\bibinfo {author} {\bibfnamefont {J.~I.}\ \bibnamefont
  {Cirac}}, \bibinfo {author} {\bibfnamefont {D.}~\bibnamefont {Perez-Garcia}},
  \bibinfo {author} {\bibfnamefont {N.}~\bibnamefont {Schuch}}, \ and\ \bibinfo
  {author} {\bibfnamefont {F.}~\bibnamefont {Verstraete}},\ }\href {\doibase
  10.1088/1742-5468/aa7e55} {\bibfield  {journal} {\bibinfo  {journal} {J.
  Stat. Mech.}\ ,\ \bibinfo {pages} {083105}} (\bibinfo {year}
  {2017})}\BibitemShut {NoStop}%
\bibitem [{\citenamefont {\c{S}ahino\v{g}lu}\ \emph {et~al.}(2018)\citenamefont
  {\c{S}ahino\v{g}lu}, \citenamefont {Shukla}, \citenamefont {Bi},\ and\
  \citenamefont {Chen}}]{sahinoglu2018matrix}%
  \BibitemOpen
  \bibfield  {author} {\bibinfo {author} {\bibfnamefont {M.~B.}\ \bibnamefont
  {\c{S}ahino\v{g}lu}}, \bibinfo {author} {\bibfnamefont {S.~K.}\ \bibnamefont
  {Shukla}}, \bibinfo {author} {\bibfnamefont {F.}~\bibnamefont {Bi}}, \ and\
  \bibinfo {author} {\bibfnamefont {X.}~\bibnamefont {Chen}},\ }\href {\doibase
  10.1103/PhysRevB.98.245122} {\bibfield  {journal} {\bibinfo  {journal} {Phys.
  Rev. B}\ }\textbf {\bibinfo {volume} {98}},\ \bibinfo {pages} {245122}
  (\bibinfo {year} {2018})}\BibitemShut {NoStop}%
\bibitem [{\citenamefont {Gong}\ \emph
  {et~al.}(2020{\natexlab{a}})\citenamefont {Gong}, \citenamefont
  {S\"underhauf}, \citenamefont {Schuch},\ and\ \citenamefont
  {Cirac}}]{gong2020classification}%
  \BibitemOpen
  \bibfield  {author} {\bibinfo {author} {\bibfnamefont {Z.}~\bibnamefont
  {Gong}}, \bibinfo {author} {\bibfnamefont {C.}~\bibnamefont {S\"underhauf}},
  \bibinfo {author} {\bibfnamefont {N.}~\bibnamefont {Schuch}}, \ and\ \bibinfo
  {author} {\bibfnamefont {J.~I.}\ \bibnamefont {Cirac}},\ }\href {\doibase
  10.1103/PhysRevLett.124.100402} {\bibfield  {journal} {\bibinfo  {journal}
  {Phys. Rev. Lett.}\ }\textbf {\bibinfo {volume} {124}},\ \bibinfo {pages}
  {100402} (\bibinfo {year} {2020}{\natexlab{a}})}\BibitemShut {NoStop}%
\bibitem [{\citenamefont {Piroli}\ \emph {et~al.}(2020)\citenamefont {Piroli},
  \citenamefont {Turzillo}, \citenamefont {Shukla},\ and\ \citenamefont
  {Cirac}}]{piroli2020fermionic}%
  \BibitemOpen
  \bibfield  {author} {\bibinfo {author} {\bibfnamefont {L.}~\bibnamefont
  {Piroli}}, \bibinfo {author} {\bibfnamefont {A.}~\bibnamefont {Turzillo}},
  \bibinfo {author} {\bibfnamefont {S.~K.}\ \bibnamefont {Shukla}}, \ and\
  \bibinfo {author} {\bibfnamefont {J.~I.}\ \bibnamefont {Cirac}},\ }\href@noop
  {} {\bibfield  {journal} {\bibinfo  {journal} {arXiv:2007.11905}\ } (\bibinfo
  {year} {2020})}\BibitemShut {NoStop}%
\bibitem [{\citenamefont {Else}\ and\ \citenamefont {Nayak}(2016)}]{Else2016}%
  \BibitemOpen
  \bibfield  {author} {\bibinfo {author} {\bibfnamefont {D.~V.}\ \bibnamefont
  {Else}}\ and\ \bibinfo {author} {\bibfnamefont {C.}~\bibnamefont {Nayak}},\
  }\href {\doibase 10.1103/PhysRevB.93.201103} {\bibfield  {journal} {\bibinfo
  {journal} {Phys. Rev. B}\ }\textbf {\bibinfo {volume} {93}},\ \bibinfo
  {pages} {201103(R)} (\bibinfo {year} {2016})}\BibitemShut {NoStop}%
\bibitem [{\citenamefont {Po}\ \emph {et~al.}(2016)\citenamefont {Po},
  \citenamefont {Fidkowski}, \citenamefont {Morimoto}, \citenamefont {Potter},\
  and\ \citenamefont {Vishwanath}}]{Po2016}%
  \BibitemOpen
  \bibfield  {author} {\bibinfo {author} {\bibfnamefont {H.~C.}\ \bibnamefont
  {Po}}, \bibinfo {author} {\bibfnamefont {L.}~\bibnamefont {Fidkowski}},
  \bibinfo {author} {\bibfnamefont {T.}~\bibnamefont {Morimoto}}, \bibinfo
  {author} {\bibfnamefont {A.~C.}\ \bibnamefont {Potter}}, \ and\ \bibinfo
  {author} {\bibfnamefont {A.}~\bibnamefont {Vishwanath}},\ }\href {\doibase
  10.1103/PhysRevX.6.041070} {\bibfield  {journal} {\bibinfo  {journal} {Phys.
  Rev. X}\ }\textbf {\bibinfo {volume} {6}},\ \bibinfo {pages} {041070}
  (\bibinfo {year} {2016})}\BibitemShut {NoStop}%
\bibitem [{\citenamefont {Potter}\ and\ \citenamefont
  {Morimoto}(2017)}]{Potter2017}%
  \BibitemOpen
  \bibfield  {author} {\bibinfo {author} {\bibfnamefont {A.~C.}\ \bibnamefont
  {Potter}}\ and\ \bibinfo {author} {\bibfnamefont {T.}~\bibnamefont
  {Morimoto}},\ }\href {\doibase 10.1103/PhysRevB.95.155126} {\bibfield
  {journal} {\bibinfo  {journal} {Phys. Rev. B}\ }\textbf {\bibinfo {volume}
  {95}},\ \bibinfo {pages} {155126} (\bibinfo {year} {2017})}\BibitemShut
  {NoStop}%
\bibitem [{\citenamefont {Harper}\ and\ \citenamefont {Roy}(2017)}]{Roy2017}%
  \BibitemOpen
  \bibfield  {author} {\bibinfo {author} {\bibfnamefont {F.}~\bibnamefont
  {Harper}}\ and\ \bibinfo {author} {\bibfnamefont {R.}~\bibnamefont {Roy}},\
  }\href {\doibase 10.1103/PhysRevLett.118.115301} {\bibfield  {journal}
  {\bibinfo  {journal} {Phys. Rev. Lett.}\ }\textbf {\bibinfo {volume} {118}},\
  \bibinfo {pages} {115301} (\bibinfo {year} {2017})}\BibitemShut {NoStop}%
\bibitem [{\citenamefont {Duschatko}\ \emph {et~al.}(2018)\citenamefont
  {Duschatko}, \citenamefont {Dumitrescu},\ and\ \citenamefont
  {Potter}}]{duschatko2018Tracking}%
  \BibitemOpen
  \bibfield  {author} {\bibinfo {author} {\bibfnamefont {B.~R.}\ \bibnamefont
  {Duschatko}}, \bibinfo {author} {\bibfnamefont {P.~T.}\ \bibnamefont
  {Dumitrescu}}, \ and\ \bibinfo {author} {\bibfnamefont {A.~C.}\ \bibnamefont
  {Potter}},\ }\href {\doibase 10.1103/PhysRevB.98.054309} {\bibfield
  {journal} {\bibinfo  {journal} {Phys. Rev. B}\ }\textbf {\bibinfo {volume}
  {98}},\ \bibinfo {pages} {054309} (\bibinfo {year} {2018})}\BibitemShut
  {NoStop}%
\bibitem [{\citenamefont {Fidkowski}\ \emph {et~al.}(2019)\citenamefont
  {Fidkowski}, \citenamefont {Po}, \citenamefont {Potter},\ and\ \citenamefont
  {Vishwanath}}]{fidkowski2019interacting}%
  \BibitemOpen
  \bibfield  {author} {\bibinfo {author} {\bibfnamefont {L.}~\bibnamefont
  {Fidkowski}}, \bibinfo {author} {\bibfnamefont {H.~C.}\ \bibnamefont {Po}},
  \bibinfo {author} {\bibfnamefont {A.~C.}\ \bibnamefont {Potter}}, \ and\
  \bibinfo {author} {\bibfnamefont {A.}~\bibnamefont {Vishwanath}},\ }\href
  {\doibase 10.1103/PhysRevB.99.085115} {\bibfield  {journal} {\bibinfo
  {journal} {Phys. Rev. B}\ }\textbf {\bibinfo {volume} {99}},\ \bibinfo
  {pages} {085115} (\bibinfo {year} {2019})}\BibitemShut {NoStop}%
\bibitem [{\citenamefont {Zhang}\ and\ \citenamefont
  {Levin}(2020)}]{zhang2020classification}%
  \BibitemOpen
  \bibfield  {author} {\bibinfo {author} {\bibfnamefont {C.}~\bibnamefont
  {Zhang}}\ and\ \bibinfo {author} {\bibfnamefont {M.}~\bibnamefont {Levin}},\
  }\href {http://arxiv.org/abs/2010.02253} {\bibfield  {journal} {\bibinfo
  {journal} {arXiv:2010.02253}\ } (\bibinfo {year} {2020})}\BibitemShut
  {NoStop}%
\bibitem [{\citenamefont {Liu}\ \emph {et~al.}(2020)\citenamefont {Liu},
  \citenamefont {Shapourian}, \citenamefont {Glorioso},\ and\ \citenamefont
  {Ryu}}]{Liu2020}%
  \BibitemOpen
  \bibfield  {author} {\bibinfo {author} {\bibfnamefont {Y.}~\bibnamefont
  {Liu}}, \bibinfo {author} {\bibfnamefont {H.}~\bibnamefont {Shapourian}},
  \bibinfo {author} {\bibfnamefont {P.}~\bibnamefont {Glorioso}}, \ and\
  \bibinfo {author} {\bibfnamefont {S.}~\bibnamefont {Ryu}},\ }\href
  {https://arxiv.org/abs/2012.08384} {\bibfield  {journal} {\bibinfo  {journal}
  {arXiv:2012.08384}\ } (\bibinfo {year} {2020})}\BibitemShut {NoStop}%
\bibitem [{\citenamefont {Shenker}\ and\ \citenamefont
  {Stanford}(2014{\natexlab{a}})}]{shenker2014multiple}%
  \BibitemOpen
  \bibfield  {author} {\bibinfo {author} {\bibfnamefont {S.~H.}\ \bibnamefont
  {Shenker}}\ and\ \bibinfo {author} {\bibfnamefont {D.}~\bibnamefont
  {Stanford}},\ }\href {\doibase 10.1007/JHEP12(2014)046} {\bibfield  {journal}
  {\bibinfo  {journal} {JHEP}\ }\textbf {\bibinfo {volume} {2014}},\ \bibinfo
  {pages} {46} (\bibinfo {year} {2014}{\natexlab{a}})}\BibitemShut {NoStop}%
\bibitem [{\citenamefont {Shenker}\ and\ \citenamefont
  {Stanford}(2014{\natexlab{b}})}]{shenker2014black}%
  \BibitemOpen
  \bibfield  {author} {\bibinfo {author} {\bibfnamefont {S.~H.}\ \bibnamefont
  {Shenker}}\ and\ \bibinfo {author} {\bibfnamefont {D.}~\bibnamefont
  {Stanford}},\ }\href {\doibase 10.1007/JHEP03(2014)067} {\bibfield  {journal}
  {\bibinfo  {journal} {JHEP}\ }\textbf {\bibinfo {volume} {2014}},\ \bibinfo
  {pages} {67} (\bibinfo {year} {2014}{\natexlab{b}})}\BibitemShut {NoStop}%
\bibitem [{\citenamefont {Roberts}\ \emph {et~al.}(2015)\citenamefont
  {Roberts}, \citenamefont {Stanford},\ and\ \citenamefont
  {Susskind}}]{roberts2015localized}%
  \BibitemOpen
  \bibfield  {author} {\bibinfo {author} {\bibfnamefont {D.~A.}\ \bibnamefont
  {Roberts}}, \bibinfo {author} {\bibfnamefont {D.}~\bibnamefont {Stanford}}, \
  and\ \bibinfo {author} {\bibfnamefont {L.}~\bibnamefont {Susskind}},\ }\href
  {\doibase 10.1007/JHEP03(2015)051} {\bibfield  {journal} {\bibinfo  {journal}
  {JHEP}\ }\textbf {\bibinfo {volume} {2015}},\ \bibinfo {pages} {51} (\bibinfo
  {year} {2015})}\BibitemShut {NoStop}%
\bibitem [{\citenamefont {Zanardi}(2001)}]{zanardi2001entanglement}%
  \BibitemOpen
  \bibfield  {author} {\bibinfo {author} {\bibfnamefont {P.}~\bibnamefont
  {Zanardi}},\ }\href {\doibase 10.1103/PhysRevA.63.040304} {\bibfield
  {journal} {\bibinfo  {journal} {Phys. Rev. A}\ }\textbf {\bibinfo {volume}
  {63}},\ \bibinfo {pages} {040304} (\bibinfo {year} {2001})}\BibitemShut
  {NoStop}%
\bibitem [{\citenamefont {Hosur}\ \emph {et~al.}(2016)\citenamefont {Hosur},
  \citenamefont {Qi}, \citenamefont {Roberts},\ and\ \citenamefont
  {Yoshida}}]{hosur2016chaos}%
  \BibitemOpen
  \bibfield  {author} {\bibinfo {author} {\bibfnamefont {P.}~\bibnamefont
  {Hosur}}, \bibinfo {author} {\bibfnamefont {X.-L.}\ \bibnamefont {Qi}},
  \bibinfo {author} {\bibfnamefont {D.~A.}\ \bibnamefont {Roberts}}, \ and\
  \bibinfo {author} {\bibfnamefont {B.}~\bibnamefont {Yoshida}},\ }\href
  {\doibase 10.1007/JHEP02(2016)004} {\bibfield  {journal} {\bibinfo  {journal}
  {JHEP}\ }\textbf {\bibinfo {volume} {2016}},\ \bibinfo {pages} {4} (\bibinfo
  {year} {2016})}\BibitemShut {NoStop}%
\bibitem [{\citenamefont {Choi}(1975)}]{Choi1975}%
  \BibitemOpen
  \bibfield  {author} {\bibinfo {author} {\bibfnamefont {M.-D.}\ \bibnamefont
  {Choi}},\ }\href {\doibase https://doi.org/10.1016/0024-3795(75)90075-0}
  {\bibfield  {journal} {\bibinfo  {journal} {Linear Alg. Appl.}\ }\textbf
  {\bibinfo {volume} {10}},\ \bibinfo {pages} {285 } (\bibinfo {year}
  {1975})}\BibitemShut {NoStop}%
\bibitem [{Note1()}]{Note1}%
  \BibitemOpen
  \bibinfo {note} {To be rigorous, we should also require $|a|+|b|\le N-2r$,
  although typically $|a|,|b|\sim \protect \mathcal {O}(r)\ll N$ is relevant to
  both practical numerical calculations and experimental measurements. The same
  constraint appears for the validity of Eq.~(\ref {eq:main_result}) and
  explains why the minimal experimental setup is $N=4$.}\BibitemShut {Stop}%
\bibitem [{SM()}]{SM}%
  \BibitemOpen
  \href@noop {} {}\bibinfo {note} {See Supplemental Material, which includes
  Ref.~\cite{Hayden2004}, for details.}\BibitemShut {Stop}%
\bibitem [{\citenamefont {Nielsen}\ and\ \citenamefont
  {Chuang}(2010)}]{Nielsen2010}%
  \BibitemOpen
  \bibfield  {author} {\bibinfo {author} {\bibfnamefont {M.~A.}\ \bibnamefont
  {Nielsen}}\ and\ \bibinfo {author} {\bibfnamefont {I.~L.}\ \bibnamefont
  {Chuang}},\ }\href@noop {} {\emph {\bibinfo {title} {Quantum Computation and
  Information}}}\ (\bibinfo  {publisher} {Cambridge University Press,
  Cambridge},\ \bibinfo {year} {2010})\BibitemShut {NoStop}%
\bibitem [{\citenamefont {Linden}\ \emph {et~al.}(2013)\citenamefont {Linden},
  \citenamefont {Mosonyi},\ and\ \citenamefont {Winter}}]{Linden2013}%
  \BibitemOpen
  \bibfield  {author} {\bibinfo {author} {\bibfnamefont {N.}~\bibnamefont
  {Linden}}, \bibinfo {author} {\bibfnamefont {M.}~\bibnamefont {Mosonyi}}, \
  and\ \bibinfo {author} {\bibfnamefont {A.}~\bibnamefont {Winter}},\ }\href
  {\doibase 10.1098/rspa.2012.0737} {\bibfield  {journal} {\bibinfo  {journal}
  {Proc. R. Soc. A}\ }\textbf {\bibinfo {volume} {469}},\ \bibinfo {pages}
  {20120737} (\bibinfo {year} {2013})}\BibitemShut {NoStop}%
\bibitem [{\citenamefont {van Dam}\ and\ \citenamefont
  {Hayden}(2002)}]{Hayden2002}%
  \BibitemOpen
  \bibfield  {author} {\bibinfo {author} {\bibfnamefont {W.}~\bibnamefont {van
  Dam}}\ and\ \bibinfo {author} {\bibfnamefont {P.}~\bibnamefont {Hayden}},\
  }\href {https://arxiv.org/abs/quant-ph/0204093} {\bibfield  {journal}
  {\bibinfo  {journal} {arXiv:quant-ph/0204093}\ } (\bibinfo {year}
  {2002})}\BibitemShut {NoStop}%
\bibitem [{\citenamefont {Affleck}\ \emph {et~al.}(1987)\citenamefont
  {Affleck}, \citenamefont {Kennedy}, \citenamefont {Lieb},\ and\ \citenamefont
  {Tasaki}}]{Tasaki1987}%
  \BibitemOpen
  \bibfield  {author} {\bibinfo {author} {\bibfnamefont {I.}~\bibnamefont
  {Affleck}}, \bibinfo {author} {\bibfnamefont {T.}~\bibnamefont {Kennedy}},
  \bibinfo {author} {\bibfnamefont {E.~H.}\ \bibnamefont {Lieb}}, \ and\
  \bibinfo {author} {\bibfnamefont {H.}~\bibnamefont {Tasaki}},\ }\href
  {\doibase 10.1103/PhysRevLett.59.799} {\bibfield  {journal} {\bibinfo
  {journal} {Phys. Rev. Lett.}\ }\textbf {\bibinfo {volume} {59}},\ \bibinfo
  {pages} {799} (\bibinfo {year} {1987})}\BibitemShut {NoStop}%
\bibitem [{\citenamefont {Zhou}\ and\ \citenamefont {Luitz}(2017)}]{Zhou2017}%
  \BibitemOpen
  \bibfield  {author} {\bibinfo {author} {\bibfnamefont {T.}~\bibnamefont
  {Zhou}}\ and\ \bibinfo {author} {\bibfnamefont {D.~J.}\ \bibnamefont
  {Luitz}},\ }\href {\doibase 10.1103/PhysRevB.95.094206} {\bibfield  {journal}
  {\bibinfo  {journal} {Phys. Rev. B}\ }\textbf {\bibinfo {volume} {95}},\
  \bibinfo {pages} {094206} (\bibinfo {year} {2017})}\BibitemShut {NoStop}%
\bibitem [{\citenamefont {Huang}(2019)}]{huang2019dynamics}%
  \BibitemOpen
  \bibfield  {author} {\bibinfo {author} {\bibfnamefont {Y.}~\bibnamefont
  {Huang}},\ }\href {https://arxiv.org/abs/1902.00977} {\bibfield  {journal}
  {\bibinfo  {journal} {arXiv:1902.00977}\ } (\bibinfo {year}
  {2019})}\BibitemShut {NoStop}%
\bibitem [{\citenamefont {Rakovszky}\ \emph {et~al.}(2019)\citenamefont
  {Rakovszky}, \citenamefont {Pollmann},\ and\ \citenamefont {von
  Keyserlingk}}]{rakovszky2019subballistic}%
  \BibitemOpen
  \bibfield  {author} {\bibinfo {author} {\bibfnamefont {T.}~\bibnamefont
  {Rakovszky}}, \bibinfo {author} {\bibfnamefont {F.}~\bibnamefont {Pollmann}},
  \ and\ \bibinfo {author} {\bibfnamefont {C.~W.}\ \bibnamefont {von
  Keyserlingk}},\ }\href {\doibase 10.1103/PhysRevLett.122.250602} {\bibfield
  {journal} {\bibinfo  {journal} {Phys. Rev. Lett.}\ }\textbf {\bibinfo
  {volume} {122}},\ \bibinfo {pages} {250602} (\bibinfo {year}
  {2019})}\BibitemShut {NoStop}%
\bibitem [{Note2()}]{Note2}%
  \BibitemOpen
  \bibinfo {note} {To avoid undesired finite-size saturation, we may choose the
  subsystem size to be no smaller than $2rt$ for a range-$r$ QCA at time $t$;
  alternatively, given subsystem size $|A|$, we should focus on a time interval
  up to $|A|/(2r)$.}\BibitemShut {Stop}%
\bibitem [{\citenamefont {Fan}\ \emph {et~al.}(2017)\citenamefont {Fan},
  \citenamefont {Zhang}, \citenamefont {Shen},\ and\ \citenamefont
  {Zhai}}]{Fan2017}%
  \BibitemOpen
  \bibfield  {author} {\bibinfo {author} {\bibfnamefont {R.}~\bibnamefont
  {Fan}}, \bibinfo {author} {\bibfnamefont {P.}~\bibnamefont {Zhang}}, \bibinfo
  {author} {\bibfnamefont {H.}~\bibnamefont {Shen}}, \ and\ \bibinfo {author}
  {\bibfnamefont {H.}~\bibnamefont {Zhai}},\ }\href {\doibase
  https://doi.org/10.1016/j.scib.2017.04.011} {\bibfield  {journal} {\bibinfo
  {journal} {Sci. Bull.}\ }\textbf {\bibinfo {volume} {62}},\ \bibinfo {pages}
  {707} (\bibinfo {year} {2017})}\BibitemShut {NoStop}%
\bibitem [{\citenamefont {Huang}\ \emph {et~al.}(2017)\citenamefont {Huang},
  \citenamefont {Zhang},\ and\ \citenamefont {Chen}}]{Chen2017}%
  \BibitemOpen
  \bibfield  {author} {\bibinfo {author} {\bibfnamefont {Y.}~\bibnamefont
  {Huang}}, \bibinfo {author} {\bibfnamefont {Y.-L.}\ \bibnamefont {Zhang}}, \
  and\ \bibinfo {author} {\bibfnamefont {X.}~\bibnamefont {Chen}},\ }\href
  {\doibase 10.1002/andp.201600318} {\bibfield  {journal} {\bibinfo  {journal}
  {Ann. Phys.}\ }\textbf {\bibinfo {volume} {529}},\ \bibinfo {pages} {1600318}
  (\bibinfo {year} {2017})}\BibitemShut {NoStop}%
\bibitem [{\citenamefont {Islam}\ \emph {et~al.}(2015)\citenamefont {Islam},
  \citenamefont {Ma}, \citenamefont {Preiss}, \citenamefont {Tai},
  \citenamefont {Lukin}, \citenamefont {Rispoli},\ and\ \citenamefont
  {Greiner}}]{Greiner2015}%
  \BibitemOpen
  \bibfield  {author} {\bibinfo {author} {\bibfnamefont {R.}~\bibnamefont
  {Islam}}, \bibinfo {author} {\bibfnamefont {R.}~\bibnamefont {Ma}}, \bibinfo
  {author} {\bibfnamefont {P.~M.}\ \bibnamefont {Preiss}}, \bibinfo {author}
  {\bibfnamefont {M.~E.}\ \bibnamefont {Tai}}, \bibinfo {author} {\bibfnamefont
  {A.}~\bibnamefont {Lukin}}, \bibinfo {author} {\bibfnamefont
  {M.}~\bibnamefont {Rispoli}}, \ and\ \bibinfo {author} {\bibfnamefont
  {M.}~\bibnamefont {Greiner}},\ }\href {http://dx.doi.org/10.1038/nature15750}
  {\bibfield  {journal} {\bibinfo  {journal} {Nature}\ }\textbf {\bibinfo
  {volume} {528}},\ \bibinfo {pages} {77} (\bibinfo {year} {2015})}\BibitemShut
  {NoStop}%
\bibitem [{\citenamefont {Kaufman}\ \emph {et~al.}(2016)\citenamefont
  {Kaufman}, \citenamefont {Tai}, \citenamefont {Lukin}, \citenamefont
  {Rispoli}, \citenamefont {Schittko}, \citenamefont {Preiss},\ and\
  \citenamefont {Greiner}}]{Greiner2016}%
  \BibitemOpen
  \bibfield  {author} {\bibinfo {author} {\bibfnamefont {A.~M.}\ \bibnamefont
  {Kaufman}}, \bibinfo {author} {\bibfnamefont {M.~E.}\ \bibnamefont {Tai}},
  \bibinfo {author} {\bibfnamefont {A.}~\bibnamefont {Lukin}}, \bibinfo
  {author} {\bibfnamefont {M.}~\bibnamefont {Rispoli}}, \bibinfo {author}
  {\bibfnamefont {R.}~\bibnamefont {Schittko}}, \bibinfo {author}
  {\bibfnamefont {P.~M.}\ \bibnamefont {Preiss}}, \ and\ \bibinfo {author}
  {\bibfnamefont {M.}~\bibnamefont {Greiner}},\ }\href {\doibase
  10.1126/science.aaf6725} {\bibfield  {journal} {\bibinfo  {journal}
  {Science}\ }\textbf {\bibinfo {volume} {353}},\ \bibinfo {pages} {794}
  (\bibinfo {year} {2016})}\BibitemShut {NoStop}%
\bibitem [{\citenamefont {Linke}\ \emph {et~al.}(2018)\citenamefont {Linke},
  \citenamefont {Johri}, \citenamefont {Figgatt}, \citenamefont {Landsman},
  \citenamefont {Matsuura},\ and\ \citenamefont {Monroe}}]{Linke2018}%
  \BibitemOpen
  \bibfield  {author} {\bibinfo {author} {\bibfnamefont {N.~M.}\ \bibnamefont
  {Linke}}, \bibinfo {author} {\bibfnamefont {S.}~\bibnamefont {Johri}},
  \bibinfo {author} {\bibfnamefont {C.}~\bibnamefont {Figgatt}}, \bibinfo
  {author} {\bibfnamefont {K.~A.}\ \bibnamefont {Landsman}}, \bibinfo {author}
  {\bibfnamefont {A.~Y.}\ \bibnamefont {Matsuura}}, \ and\ \bibinfo {author}
  {\bibfnamefont {C.}~\bibnamefont {Monroe}},\ }\href {\doibase
  10.1103/PhysRevA.98.052334} {\bibfield  {journal} {\bibinfo  {journal} {Phys.
  Rev. A}\ }\textbf {\bibinfo {volume} {98}},\ \bibinfo {pages} {052334}
  (\bibinfo {year} {2018})}\BibitemShut {NoStop}%
\bibitem [{\citenamefont {Lukin}\ \emph {et~al.}(2019)\citenamefont {Lukin},
  \citenamefont {Rispoli}, \citenamefont {Schittko}, \citenamefont {Tai},
  \citenamefont {Kaufman}, \citenamefont {Choi}, \citenamefont {Khemani},
  \citenamefont {L{\'e}onard},\ and\ \citenamefont {Greiner}}]{Greiner2019}%
  \BibitemOpen
  \bibfield  {author} {\bibinfo {author} {\bibfnamefont {A.}~\bibnamefont
  {Lukin}}, \bibinfo {author} {\bibfnamefont {M.}~\bibnamefont {Rispoli}},
  \bibinfo {author} {\bibfnamefont {R.}~\bibnamefont {Schittko}}, \bibinfo
  {author} {\bibfnamefont {M.~E.}\ \bibnamefont {Tai}}, \bibinfo {author}
  {\bibfnamefont {A.~M.}\ \bibnamefont {Kaufman}}, \bibinfo {author}
  {\bibfnamefont {S.}~\bibnamefont {Choi}}, \bibinfo {author} {\bibfnamefont
  {V.}~\bibnamefont {Khemani}}, \bibinfo {author} {\bibfnamefont
  {J.}~\bibnamefont {L{\'e}onard}}, \ and\ \bibinfo {author} {\bibfnamefont
  {M.}~\bibnamefont {Greiner}},\ }\href {\doibase 10.1126/science.aau0818}
  {\bibfield  {journal} {\bibinfo  {journal} {Science}\ }\textbf {\bibinfo
  {volume} {364}},\ \bibinfo {pages} {256} (\bibinfo {year}
  {2019})}\BibitemShut {NoStop}%
\bibitem [{\citenamefont {Horodecki}\ and\ \citenamefont
  {Ekert}(2002)}]{Ekert2002}%
  \BibitemOpen
  \bibfield  {author} {\bibinfo {author} {\bibfnamefont {P.}~\bibnamefont
  {Horodecki}}\ and\ \bibinfo {author} {\bibfnamefont {A.}~\bibnamefont
  {Ekert}},\ }\href {\doibase 10.1103/PhysRevLett.89.127902} {\bibfield
  {journal} {\bibinfo  {journal} {Phys. Rev. Lett.}\ }\textbf {\bibinfo
  {volume} {89}},\ \bibinfo {pages} {127902} (\bibinfo {year}
  {2002})}\BibitemShut {NoStop}%
\bibitem [{\citenamefont {Abanin}\ and\ \citenamefont
  {Demler}(2012)}]{Abanin2012}%
  \BibitemOpen
  \bibfield  {author} {\bibinfo {author} {\bibfnamefont {D.~A.}\ \bibnamefont
  {Abanin}}\ and\ \bibinfo {author} {\bibfnamefont {E.}~\bibnamefont
  {Demler}},\ }\href {\doibase 10.1103/PhysRevLett.109.020504} {\bibfield
  {journal} {\bibinfo  {journal} {Phys. Rev. Lett.}\ }\textbf {\bibinfo
  {volume} {109}},\ \bibinfo {pages} {020504} (\bibinfo {year}
  {2012})}\BibitemShut {NoStop}%
\bibitem [{\citenamefont {Daley}\ \emph {et~al.}(2012)\citenamefont {Daley},
  \citenamefont {Pichler}, \citenamefont {Schachenmayer},\ and\ \citenamefont
  {Zoller}}]{Daley2012}%
  \BibitemOpen
  \bibfield  {author} {\bibinfo {author} {\bibfnamefont {A.~J.}\ \bibnamefont
  {Daley}}, \bibinfo {author} {\bibfnamefont {H.}~\bibnamefont {Pichler}},
  \bibinfo {author} {\bibfnamefont {J.}~\bibnamefont {Schachenmayer}}, \ and\
  \bibinfo {author} {\bibfnamefont {P.}~\bibnamefont {Zoller}},\ }\href
  {\doibase 10.1103/PhysRevLett.109.020505} {\bibfield  {journal} {\bibinfo
  {journal} {Phys. Rev. Lett.}\ }\textbf {\bibinfo {volume} {109}},\ \bibinfo
  {pages} {020505} (\bibinfo {year} {2012})}\BibitemShut {NoStop}%
\bibitem [{\citenamefont {Blatt}\ and\ \citenamefont {Roos}(2012)}]{Blatt2012}%
  \BibitemOpen
  \bibfield  {author} {\bibinfo {author} {\bibfnamefont {R.}~\bibnamefont
  {Blatt}}\ and\ \bibinfo {author} {\bibfnamefont {C.~F.}\ \bibnamefont
  {Roos}},\ }\href {\doibase 10.1038/nphys2252} {\bibfield  {journal} {\bibinfo
   {journal} {Nat. Phys.}\ }\textbf {\bibinfo {volume} {8}},\ \bibinfo {pages}
  {277} (\bibinfo {year} {2012})}\BibitemShut {NoStop}%
\bibitem [{\citenamefont {Houck}\ \emph {et~al.}(2012)\citenamefont {Houck},
  \citenamefont {T\"ureci},\ and\ \citenamefont {Koch}}]{Koch2012}%
  \BibitemOpen
  \bibfield  {author} {\bibinfo {author} {\bibfnamefont {A.~A.}\ \bibnamefont
  {Houck}}, \bibinfo {author} {\bibfnamefont {H.~E.}\ \bibnamefont {T\"ureci}},
  \ and\ \bibinfo {author} {\bibfnamefont {J.}~\bibnamefont {Koch}},\ }\href
  {\doibase 10.1038/nphys2251} {\bibfield  {journal} {\bibinfo  {journal} {Nat.
  Phys.}\ }\textbf {\bibinfo {volume} {8}},\ \bibinfo {pages} {292} (\bibinfo
  {year} {2012})}\BibitemShut {NoStop}%
\bibitem [{\citenamefont {Levine}\ \emph {et~al.}(2018)\citenamefont {Levine},
  \citenamefont {Keesling}, \citenamefont {Omran}, \citenamefont {Bernien},
  \citenamefont {Schwartz}, \citenamefont {Zibrov}, \citenamefont {Endres},
  \citenamefont {Greiner}, \citenamefont {Vuleti\ifmmode~\acute{c}\else
  \'{c}\fi{}},\ and\ \citenamefont {Lukin}}]{Levine2018}%
  \BibitemOpen
  \bibfield  {author} {\bibinfo {author} {\bibfnamefont {H.}~\bibnamefont
  {Levine}}, \bibinfo {author} {\bibfnamefont {A.}~\bibnamefont {Keesling}},
  \bibinfo {author} {\bibfnamefont {A.}~\bibnamefont {Omran}}, \bibinfo
  {author} {\bibfnamefont {H.}~\bibnamefont {Bernien}}, \bibinfo {author}
  {\bibfnamefont {S.}~\bibnamefont {Schwartz}}, \bibinfo {author}
  {\bibfnamefont {A.~S.}\ \bibnamefont {Zibrov}}, \bibinfo {author}
  {\bibfnamefont {M.}~\bibnamefont {Endres}}, \bibinfo {author} {\bibfnamefont
  {M.}~\bibnamefont {Greiner}}, \bibinfo {author} {\bibfnamefont
  {V.}~\bibnamefont {Vuleti\ifmmode~\acute{c}\else \'{c}\fi{}}}, \ and\
  \bibinfo {author} {\bibfnamefont {M.~D.}\ \bibnamefont {Lukin}},\ }\href
  {\doibase 10.1103/PhysRevLett.121.123603} {\bibfield  {journal} {\bibinfo
  {journal} {Phys. Rev. Lett.}\ }\textbf {\bibinfo {volume} {121}},\ \bibinfo
  {pages} {123603} (\bibinfo {year} {2018})}\BibitemShut {NoStop}%
\bibitem [{\citenamefont {Elben}\ \emph {et~al.}(2018)\citenamefont {Elben},
  \citenamefont {Vermersch}, \citenamefont {Dalmonte}, \citenamefont {Cirac},\
  and\ \citenamefont {Zoller}}]{Elben2018}%
  \BibitemOpen
  \bibfield  {author} {\bibinfo {author} {\bibfnamefont {A.}~\bibnamefont
  {Elben}}, \bibinfo {author} {\bibfnamefont {B.}~\bibnamefont {Vermersch}},
  \bibinfo {author} {\bibfnamefont {M.}~\bibnamefont {Dalmonte}}, \bibinfo
  {author} {\bibfnamefont {J.~I.}\ \bibnamefont {Cirac}}, \ and\ \bibinfo
  {author} {\bibfnamefont {P.}~\bibnamefont {Zoller}},\ }\href {\doibase
  10.1103/PhysRevLett.120.050406} {\bibfield  {journal} {\bibinfo  {journal}
  {Phys. Rev. Lett.}\ }\textbf {\bibinfo {volume} {120}},\ \bibinfo {pages}
  {050406} (\bibinfo {year} {2018})}\BibitemShut {NoStop}%
\bibitem [{\citenamefont {Vermersch}\ \emph {et~al.}(2018)\citenamefont
  {Vermersch}, \citenamefont {Elben}, \citenamefont {Dalmonte}, \citenamefont
  {Cirac},\ and\ \citenamefont {Zoller}}]{Vermersch2018}%
  \BibitemOpen
  \bibfield  {author} {\bibinfo {author} {\bibfnamefont {B.}~\bibnamefont
  {Vermersch}}, \bibinfo {author} {\bibfnamefont {A.}~\bibnamefont {Elben}},
  \bibinfo {author} {\bibfnamefont {M.}~\bibnamefont {Dalmonte}}, \bibinfo
  {author} {\bibfnamefont {J.~I.}\ \bibnamefont {Cirac}}, \ and\ \bibinfo
  {author} {\bibfnamefont {P.}~\bibnamefont {Zoller}},\ }\href {\doibase
  10.1103/PhysRevA.97.023604} {\bibfield  {journal} {\bibinfo  {journal} {Phys.
  Rev. A}\ }\textbf {\bibinfo {volume} {97}},\ \bibinfo {pages} {023604}
  (\bibinfo {year} {2018})}\BibitemShut {NoStop}%
\bibitem [{\citenamefont {Brydges}\ \emph {et~al.}(2019)\citenamefont
  {Brydges}, \citenamefont {Elben}, \citenamefont {Jurcevic}, \citenamefont
  {Vermersch}, \citenamefont {Maier}, \citenamefont {Lanyon}, \citenamefont
  {Zoller}, \citenamefont {Blatt},\ and\ \citenamefont {Roos}}]{Roos2019}%
  \BibitemOpen
  \bibfield  {author} {\bibinfo {author} {\bibfnamefont {T.}~\bibnamefont
  {Brydges}}, \bibinfo {author} {\bibfnamefont {A.}~\bibnamefont {Elben}},
  \bibinfo {author} {\bibfnamefont {P.}~\bibnamefont {Jurcevic}}, \bibinfo
  {author} {\bibfnamefont {B.}~\bibnamefont {Vermersch}}, \bibinfo {author}
  {\bibfnamefont {C.}~\bibnamefont {Maier}}, \bibinfo {author} {\bibfnamefont
  {B.~P.}\ \bibnamefont {Lanyon}}, \bibinfo {author} {\bibfnamefont
  {P.}~\bibnamefont {Zoller}}, \bibinfo {author} {\bibfnamefont
  {R.}~\bibnamefont {Blatt}}, \ and\ \bibinfo {author} {\bibfnamefont {C.~F.}\
  \bibnamefont {Roos}},\ }\href {\doibase 10.1126/science.aau4963} {\bibfield
  {journal} {\bibinfo  {journal} {Science}\ }\textbf {\bibinfo {volume}
  {364}},\ \bibinfo {pages} {260} (\bibinfo {year} {2019})}\BibitemShut
  {NoStop}%
\bibitem [{\citenamefont {Osborne}(2006)}]{Osborne2006}%
  \BibitemOpen
  \bibfield  {author} {\bibinfo {author} {\bibfnamefont {T.~J.}\ \bibnamefont
  {Osborne}},\ }\href {\doibase 10.1103/PhysRevLett.97.157202} {\bibfield
  {journal} {\bibinfo  {journal} {Phys. Rev. Lett.}\ }\textbf {\bibinfo
  {volume} {97}},\ \bibinfo {pages} {157202} (\bibinfo {year}
  {2006})}\BibitemShut {NoStop}%
\bibitem [{\citenamefont {Lieb}\ and\ \citenamefont
  {Robinson}(1972)}]{Lieb1972}%
  \BibitemOpen
  \bibfield  {author} {\bibinfo {author} {\bibfnamefont {E.~H.}\ \bibnamefont
  {Lieb}}\ and\ \bibinfo {author} {\bibfnamefont {D.~W.}\ \bibnamefont
  {Robinson}},\ }\href {https://doi.org/10.1007/978-3-662-10018-9_25}
  {\bibfield  {journal} {\bibinfo  {journal} {Commun. Math. Phys.}\ }\textbf
  {\bibinfo {volume} {28}},\ \bibinfo {pages} {251} (\bibinfo {year}
  {1972})}\BibitemShut {NoStop}%
\bibitem [{\citenamefont {Bravyi}\ \emph {et~al.}(2006)\citenamefont {Bravyi},
  \citenamefont {Hastings},\ and\ \citenamefont {Verstraete}}]{Bravyi2006}%
  \BibitemOpen
  \bibfield  {author} {\bibinfo {author} {\bibfnamefont {S.}~\bibnamefont
  {Bravyi}}, \bibinfo {author} {\bibfnamefont {M.~B.}\ \bibnamefont
  {Hastings}}, \ and\ \bibinfo {author} {\bibfnamefont {F.}~\bibnamefont
  {Verstraete}},\ }\href {\doibase 10.1103/PhysRevLett.97.050401} {\bibfield
  {journal} {\bibinfo  {journal} {Phys. Rev. Lett.}\ }\textbf {\bibinfo
  {volume} {97}},\ \bibinfo {pages} {050401} (\bibinfo {year}
  {2006})}\BibitemShut {NoStop}%
\bibitem [{\citenamefont {Nachtergaele}\ and\ \citenamefont
  {Sims}(2006)}]{Nachtergaele2006}%
  \BibitemOpen
  \bibfield  {author} {\bibinfo {author} {\bibfnamefont {B.}~\bibnamefont
  {Nachtergaele}}\ and\ \bibinfo {author} {\bibfnamefont {R.}~\bibnamefont
  {Sims}},\ }\href {\doibase 10.1007/s00220-006-1556-1} {\bibfield  {journal}
  {\bibinfo  {journal} {Commun. Math. Phys.}\ }\textbf {\bibinfo {volume}
  {265}},\ \bibinfo {pages} {119} (\bibinfo {year} {2006})}\BibitemShut
  {NoStop}%
\bibitem [{Note3()}]{Note3}%
  \BibitemOpen
  \bibinfo {note} {Very recently, this is partially solved in Ref.~\cite
  {Ranard2020} for infinite and finite open chains, but remains unsolved for
  finite periodic systems discussed here.}\BibitemShut {Stop}%
\bibitem [{\citenamefont {Hastings}(2010)}]{Hastings2010}%
  \BibitemOpen
  \bibfield  {author} {\bibinfo {author} {\bibfnamefont {M.~B.}\ \bibnamefont
  {Hastings}},\ }\href@noop {} {\enquote {\bibinfo {title} {Locality in quantum
  systems},}\ } (\bibinfo {year} {2010}),\ \bibinfo {note}
  {arXiv:1008.5137}\BibitemShut {NoStop}%
\bibitem [{\citenamefont {Gong}\ \emph
  {et~al.}(2020{\natexlab{b}})\citenamefont {Gong}, \citenamefont {Yoshioka},
  \citenamefont {Shibata},\ and\ \citenamefont {Hamazaki}}]{Gong2020}%
  \BibitemOpen
  \bibfield  {author} {\bibinfo {author} {\bibfnamefont {Z.}~\bibnamefont
  {Gong}}, \bibinfo {author} {\bibfnamefont {N.}~\bibnamefont {Yoshioka}},
  \bibinfo {author} {\bibfnamefont {N.}~\bibnamefont {Shibata}}, \ and\
  \bibinfo {author} {\bibfnamefont {R.}~\bibnamefont {Hamazaki}},\ }\href
  {\doibase 10.1103/PhysRevA.101.052122} {\bibfield  {journal} {\bibinfo
  {journal} {Phys. Rev. A}\ }\textbf {\bibinfo {volume} {101}},\ \bibinfo
  {pages} {052122} (\bibinfo {year} {2020}{\natexlab{b}})}\BibitemShut
  {NoStop}%
\bibitem [{\citenamefont {Maldacena}\ \emph {et~al.}(2016)\citenamefont
  {Maldacena}, \citenamefont {Shenker},\ and\ \citenamefont
  {Stanford}}]{Stanford2016}%
  \BibitemOpen
  \bibfield  {author} {\bibinfo {author} {\bibfnamefont {J.}~\bibnamefont
  {Maldacena}}, \bibinfo {author} {\bibfnamefont {S.~H.}\ \bibnamefont
  {Shenker}}, \ and\ \bibinfo {author} {\bibfnamefont {D.}~\bibnamefont
  {Stanford}},\ }\href {\doibase 10.1007/JHEP08(2016)106} {\bibfield  {journal}
  {\bibinfo  {journal} {JHEP}\ }\textbf {\bibinfo {volume} {2016}},\ \bibinfo
  {pages} {106} (\bibinfo {year} {2016})}\BibitemShut {NoStop}%
\bibitem [{\citenamefont {Ranard}\ \emph {et~al.}(2020)\citenamefont {Ranard},
  \citenamefont {Walter},\ and\ \citenamefont {Witteveen}}]{Ranard2020}%
  \BibitemOpen
  \bibfield  {author} {\bibinfo {author} {\bibfnamefont {D.}~\bibnamefont
  {Ranard}}, \bibinfo {author} {\bibfnamefont {M.}~\bibnamefont {Walter}}, \
  and\ \bibinfo {author} {\bibfnamefont {F.}~\bibnamefont {Witteveen}},\ }\href
  {http://arxiv.org/abs/2012.00741} {\bibfield  {journal} {\bibinfo  {journal}
  {arXiv:2012.00741}\ } (\bibinfo {year} {2020})}\BibitemShut {NoStop}%
\bibitem [{\citenamefont {Hayden}\ \emph {et~al.}(2004)\citenamefont {Hayden},
  \citenamefont {Jozsa}, \citenamefont {Petz},\ and\ \citenamefont
  {Winter}}]{Hayden2004}%
  \BibitemOpen
  \bibfield  {author} {\bibinfo {author} {\bibfnamefont {P.}~\bibnamefont
  {Hayden}}, \bibinfo {author} {\bibfnamefont {R.}~\bibnamefont {Jozsa}},
  \bibinfo {author} {\bibfnamefont {D.}~\bibnamefont {Petz}}, \ and\ \bibinfo
  {author} {\bibfnamefont {A.}~\bibnamefont {Winter}},\ }\href
  {https://doi.org/10.1007/s00220-004-1049-z} {\bibfield  {journal} {\bibinfo
  {journal} {Commun. Math. Phys.}\ }\textbf {\bibinfo {volume} {246}},\
  \bibinfo {pages} {359} (\bibinfo {year} {2004})}\BibitemShut {NoStop}%
\bibitem [{Note4()}]{Note4}%
  \BibitemOpen
  \bibinfo {note} {This input should nevertheless be short-range correlated.
  Otherwise, it may mediate a nonzero correlation between $a'_{l,r}$ and
  $s'_{r,l}$ and thus Eq.~(\ref {fac}) may break down.}\BibitemShut {Stop}%
\end{thebibliography}%

\clearpage
\begin{center}
\textbf{\large Supplemental Materials}
\end{center}
\setcounter{equation}{0}
\setcounter{figure}{0}
\setcounter{table}{0}
\makeatletter
\renewcommand{\theequation}{S\arabic{equation}}
\renewcommand{\thefigure}{S\arabic{figure}}
\renewcommand{\bibnumfmt}[1]{[S#1]}

We provide some technical details supporting the claims in the main text. In particular, we discuss the validity of the bilayer representation of QCAs, and present a direct proof that the expression in Eq.~\eqref{dS} does not depend on the position where it is computed. Furthermore, we show the equivalence between Eq.~\eqref{dS} and other reformulations of the index in literature, and also report a proof of the stability of the lower bound against ``exponential tails''.

\section{Structure theorem for 1D quantum cellular automata}
Let us sketch some important results in Ref.~\cite{gross2012index} which are necessary for deriving the representation in Fig.~\ref{fig2}(a). Following the notation of Ref.~\cite{gross2012index}, we denote the action of a unitary QCA $U$ as $\alpha(\cdot)=U\cdot U^\dag$. Assuming that $\alpha$ has range $r$, then blocking at least $r$ sites into one makes it a range-1 QCA:
\begin{equation}
\alpha(\mathcal{A}_x)\subset \mathcal{A}_{x-1}\otimes \mathcal{A}_x\otimes \mathcal{A}_{x+1},
\end{equation}
where $\mathcal{A}_x$ denotes the algebra generated by the operators supported on $x$. We set
\begin{equation}
\begin{split}
\mathcal{R}_{2x}&\equiv \mathcal{S}(\alpha (\mathcal{A}_{2x}\otimes \mathcal{A}_{2x+1}),\mathcal{A}_{2x-1}\otimes\mathcal{A}_{2x}), \\
\mathcal{R}_{2x+1}&\equiv \mathcal{S}(\alpha (\mathcal{A}_{2x}\otimes \mathcal{A}_{2x+1}),\mathcal{A}_{2x+1}\otimes\mathcal{A}_{2x+2}), 
\end{split}
\end{equation}
where $S(\mathcal{A},\mathcal{B})$ denotes the support algebra of $\mathcal{A}$ on $\mathcal{B}$, which is defined as the minimal algebra $\mathcal{C}$ such that $\mathcal{A}\subseteq\mathcal{C}\otimes \mathcal{B}$. Then, we have
\begin{equation}
\begin{split}
\alpha(\mathcal{A}_{2x}\otimes\mathcal{A}_{2x+1})&\subseteq\mathcal{R}_{2x}\otimes\mathcal{R}_{2x+1},\\
\mathcal{R}_{2x+1}\otimes\mathcal{R}_{2x+2}&\subseteq\mathcal{A}_{2x+1}\otimes\mathcal{A}_{2x+2}.
\end{split}
\end{equation}
Here we have used the fact that $\mathcal{R}_x$ commutes with $\mathcal{R}_y$ $\forall x\neq y$~\cite{gross2012index}. Since $\alpha$ is an automorphism on $\bigotimes^{2l}_{x=1}\mathcal{A}_x$, this implies
\begin{equation}
\begin{split}
\alpha(\mathcal{A}_{2x}\otimes\mathcal{A}_{2x+1})&=\mathcal{R}_{2x}\otimes\mathcal{R}_{2x+1},\\
\mathcal{R}_{2x+1}\otimes\mathcal{R}_{2x+2}&=\mathcal{A}_{2x+1}\otimes\mathcal{A}_{2x+2}.
\end{split}
\end{equation}
Moreover, one can show that $\mathcal{R}_x$ should be simple (i.e., with a trivial center) and can thus be associated with a Hilbert space, analogously to $\mathcal{A}_x$. In turn, this implies that the QCA admits the following representation [cf. also Fig.~\ref{fig2}(a) in the main text]:
\begin{equation}
\begin{tikzpicture}
\Text[x=-0.7,y=0.3]{$v$}
\Text[x=-0.7,y=-0.3]{$u$}
\foreach \x in {0,...,4}
{\draw[thick,fill=yellow!20!white] (\x-0.4,-0.5) rectangle (\x+0.4,-0.1);
\draw[ultra thick] (\x-0.25,-0.1) -- (\x-0.25,0.1);
\draw[ultra thick,dotted] (\x+0.25,-0.1) -- (\x+0.25,0.1);
\draw[thick] (\x-0.25,-0.5) -- (\x-0.25,-0.7) (\x+0.25,-0.5) -- (\x+0.25,-0.7);}
\foreach \x in {1,...,4}
{\draw[thick,fill=orange!20!white] (\x-0.9,0.5) rectangle (\x-0.1,0.1);
\draw[thick] (\x-0.75,0.5) -- (\x-0.75,0.7) (\x-0.25,0.5) -- (\x-0.25,0.7);}
\fill[orange!20!white] (-0.5,0.1) rectangle (-0.1,0.5);
\draw[thick] (-0.5,0.1) -- (-0.1,0.1) -- (-0.1,0.5) -- (-0.5,0.5);
\draw[thick] (-0.25,0.5) -- (-0.25,0.7);
\fill[orange!20!white] (4.5,0.1) rectangle (4.1,0.5);
\draw[thick] (4.5,0.5) -- (4.1,0.5) -- (4.1,0.1) -- (4.5,0.1);
\draw[thick] (4.25,0.5) -- (4.25,0.7);
\end{tikzpicture},
\label{SMbl}
\end{equation}
where the upper and lower legs correspond to the (blocked) local physical Hilbert spaces where $\mathcal{A}_x$ act on, while the thick and dotted legs in the middle correspond to the virtual Hilbert spaces where $\mathcal{R}_{2x}$ and $\mathcal{R}_{2x+1}$ act on, respectively. This representation is consistent with the ``standard form'' of matrix product unitaries derived in Ref.~\cite{cirac2017matrix}, although here the QCA, as well as the blocking, need not to be homogeneous. For this reason, we have and explicit dependence on $x$ for $u_{2x,2x+1}$ and $v_{2x-1,2x}$. Note also that even the physical dimensions can be different due to an inhomogeneous blocking.

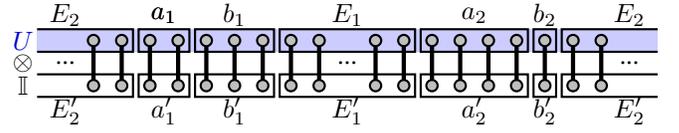
\begin{figure}[!t]
\begin{center}
\begin{tikzpicture}[scale=0.75]
      \fill[blue!20!white] (-1,0.2) rectangle (10,0.6); 
       \foreach \x in {0,...,8}
       {
           \draw[ultra thick] (0.5*\x,0.4) -- (0.5*\x,-0.4);
           \draw[thick,fill=gray!50!white] (0.5*\x,0.4) circle (0.1) (0.5*\x,-0.4) circle (0.1);       
       }
       \foreach \x in {10,...,18}
       {
           \draw[ultra thick] (0.5*\x,0.4) -- (0.5*\x,-0.4);
           \draw[thick,fill=gray!50!white] (0.5*\x,0.4) circle (0.1) (0.5*\x,-0.4) circle (0.1);         
       }
       \Text[x=-1.25,y=0.4]{$\textcolor{blue!80!black}{U}$}
       \Text[x=-1.25,y=0]{$\otimes$}
       \Text[x=-1.25,y=-0.4]{$\mathbb{I}$}
       \Text[x=4.5,y=0]{...}
       \Text[x=9.5,y=0]{...}
       \Text[x=-0.5,y=0]{...}
       \Text[x=1.25,y=0.85,]{$a_1$}
       \draw[thick] (0.8,0.2) rectangle (1.7,0.6); \Text[x=1.25,y=0.85]{$a_1$} 
       \draw[thick] (0.8,-0.2) rectangle (1.7,-0.6); \Text[x=1.25,y=-0.85]{$a'_1$} 
       \draw[thick] (1.8,0.2) rectangle (3.2,0.6); \Text[x=2.5,y=0.85]{$b_1$} 
       \draw[thick] (1.8,-0.2) rectangle (3.2,-0.6); \Text[x=2.5,y=-0.85]{$b'_1$} 
       \draw[thick] (3.3,0.2) rectangle (5.7,0.6); \Text[x=4.5,y=0.85]{$E_1$} 
       \draw[thick] (3.3,-0.2) rectangle (5.7,-0.6); \Text[x=4.5,y=-0.85]{$E'_1$} 
       \draw[thick] (5.8,0.2) rectangle (7.7,0.6); \Text[x=6.75,y=0.85]{$a_2$} 
       \draw[thick] (5.8,-0.2) rectangle (7.7,-0.6); \Text[x=6.75,y=-0.85]{$a'_2$} 
       \draw[thick] (7.8,0.2) rectangle (8.2,0.6); \Text[x=8,y=0.85]{$b_2$} 
       \draw[thick] (7.8,-0.2) rectangle (8.2,-0.6); \Text[x=8,y=-0.85]{$b'_2$} 
       \draw[thick] (10,0.6) -- (8.3,0.6) -- (8.3,0.2) -- (10,0.2) (-1,0.6) -- (0.7,0.6) -- (0.7,0.2) -- (-1,0.2); \Text[x=9.5,y=0.85]{$E_2$} \Text[x=-0.5,y=0.85]{$E_2$} 
       \draw[thick] (10,-0.6) -- (8.3,-0.6) -- (8.3,-0.2) -- (10,-0.2) (-1,-0.6) -- (0.7,-0.6) -- (0.7,-0.2) -- (-1,-0.2); \Text[x=9.5,y=-0.85]{$E'_2$} \Text[x=-0.5,y=-0.85]{$E'_2$} 
\end{tikzpicture}
\end{center}
   \caption{Subsystems of the CJS $|U\rangle\equiv(U\otimes \mathbb{I})|I\rangle$ relevant to the proof of position independence.}
      \label{figS}
\end{figure}

\section{Position independence of the entropy difference}
We prove that the entropy difference in Eq.~(\ref{dS}) is globally well-defined, in the sense that it does not depend on how large (if not too small) and where $a$, $b$ are. To this end, we only have to prove
\begin{equation}
S_\alpha(\rho_{a_1b'_1a'_2b_2})=S_\alpha(\rho_{a'_1b_1a_2b'_2}),
\label{LRLR}
\end{equation}
where $a_1$ ($b_1$) is not necessarily the same as $a_2$ ($b_2$) and $a_1\sqcup b_1$ is well separated from $a_2\sqcup b_2$. Indeed, using
\begin{equation}
\begin{split}
\rho_{a_1b'_1a'_2b_2}&=\rho_{a_1b'_1}\otimes \rho_{a'_2b_2},\\
\rho_{a'_1b_1a_2b'_2}&=\rho_{a'_1b_1}\otimes \rho_{a_2b'_2},
\end{split}
\end{equation}
which follows from the locality properties of QCA~\cite{piroli2020quantum}, we see that Eq.~\eqref{LRLR} implies $S_\alpha(\rho_{a_1b'_1})+S_\alpha(\rho_{a'_2b_2})=S_\alpha(\rho_{a'_1b_1})+S_\alpha(\rho_{a_2b'_2})$, namely
\begin{equation}
\begin{split}
S_\alpha(\rho_{a_1b'_1})-S_\alpha(\rho_{a'_1b_1})=S_\alpha(\rho_{a_2b'_2})-S_\alpha(\rho_{a'_2b_2}).
\end{split}
\end{equation}
Note that showing the equality for two well separated segments already implies the equality everywhere. This is because for two nearby segments $a_1b_1$ and $a_2b_2$ we can always find a third segment $a_3b_3$ far away from both, so that $S_\alpha(\rho_{a_1b'_1})-S_\alpha(\rho_{a'_1b_1})=S_\alpha(\rho_{a_3b'_3})-S_\alpha(\rho_{a'_3b_3})=S_\alpha(\rho_{a_2b'_2})-S_\alpha(\rho_{a'_2b_2})$. 

Let us return to the proof of Eq.~(\ref{LRLR}). Defining the regions sandwiched by $b_{1,2}$ and $a_{2,1}$ as $E_{1,2}$ [cf. Fig.~\ref{figS}], we have a bipartition $a_1E_2b_2 a'_2 E'_1 b'_1\sqcup a'_1E'_2b'_2 a_2 E_1 b_1$ (as usual, primed letters correspond to ancillary regions), and thus  
\begin{equation}
S_\alpha(\rho_{a_1E_2b_2 a'_2 E'_1 b'_1})=S_\alpha(\rho_{a'_1E'_2b'_2 a_2 E_1 b_1}).
\label{LRELRE}
\end{equation}
Now, using one again the locality properties of QCAs~\cite{piroli2020quantum}, we have
\begin{equation}
\rho_{a_1E_2b_2 a'_2 E'_1 b'_1}=\rho_{a_1b_2 a'_2 b'_1}\otimes \frac{\mathbb{I}_{E_2E'_1}}{d_{E_1}d_{E_2}},
\label{d1}
\end{equation}
and
\begin{equation}
\rho_{a'_1E'_2b'_2 a_2 E_1 b_1}=\rho_{a'_1b'_2 a_2 b_1}\otimes \frac{\mathbb{I}_{E'_2E_1}}{d_{E_1}d_{E_2}}.
\label{d1p}
\end{equation}
Combining Eqs.~(\ref{LRELRE}), (\ref{d1}) and (\ref{d1p}), we finally end up with Eq~(\ref{LRLR}).


Having in mind the position independence of Eq.~(\ref{dS}), we can directly understand why this is a topological invariant without knowing its equivalence to the index. We exploit a similar argument in Ref.~\cite{duschatko2018Tracking} and recall that a continuous deformation of QCA is nothing but a composition with a quantum circuit consisting of local unitary gates. For a given deformation, we can always apply these
local gates sequentially such that each time there is always a region on which no gates act. Then the entropy difference stays unchanged on that region, so should all the entropy differences.

\section{Equivalence to some quantities in the literature}
\subsection{Equivalence to the chiral mutual information}
\begin{figure}[!t]
\begin{center}
\includegraphics[width=8.5cm, clip]{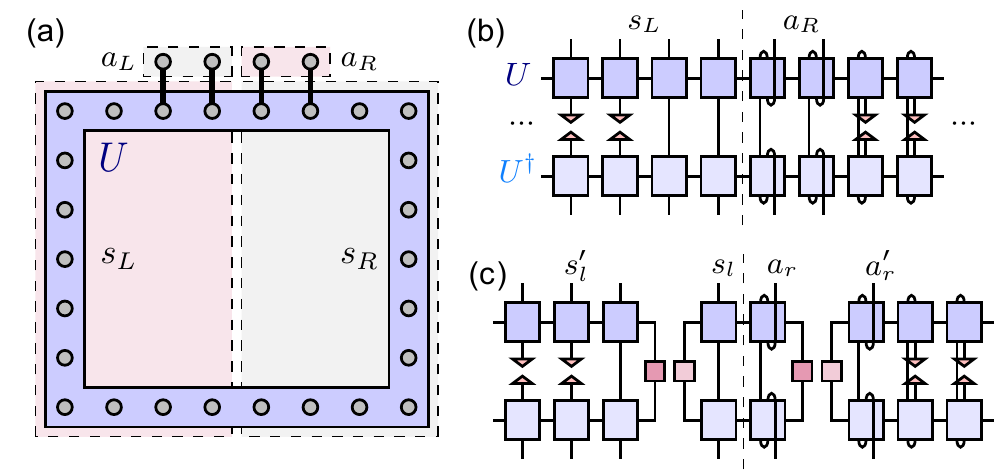}
\end{center}
   \caption{(a) ``Local CJS" (\ref{LCJS}) and its multi-partitions used to define the CMI (\ref{CMI}). See also Fig.~1 in Ref.~\cite{duschatko2018Tracking}. (b) Reduced state on $s_La_R$ with the QCA represented as a matrix-product unitary. 
   (c) The simple property \cite{cirac2017matrix,piroli2020quantum} allows us to factorize the reduced state following Eq.~(\ref{fac0}). Note that $\rho_{s_la_r}$ does not depend on the input product states (triangles).}
      \label{figS3}
\end{figure}

In this section we discuss the equivalence between Eq.~\eqref{dS} and the chiral mutual information (CMI) $\chi$ introduced in Ref.~\cite{duschatko2018Tracking}. This quantity is defined in terms of a ``local CJS" with respect to a range-$r$ QCA $U$:
\begin{equation}
|U;a\rangle\equiv (U\otimes\openone_a) |\psi_{\rm prod}\rangle \otimes  |I_a\rangle,
\label{LCJS}
\end{equation}
where $|I_a\rangle$ is the maximally entangled state between a local ancilla region $a$ and the corresponding region in the system, and $|\psi_{\rm prod}\rangle$ can be an arbitrary product state on the remaining system. Given a system bipartition $s_L\sqcup s_R$, we choose $a$ such that it is located across one of the entanglement cuts and is thus divided into $a_L$ and $a_R$,  cf. Fig.~\ref{figS3}(a). Denoting the quantum mutual information by $I(A:B)\equiv S(\rho_A)+S(\rho_B)-S(\rho_{AB})$, where $\rho_A$ is the reduced state of $|U;a\rangle$ on $A$, we define $\chi$ as~\cite{duschatko2018Tracking} 
\begin{equation}
\chi\equiv\frac{1}{2}(I(a_L: s_R) - I (a_R: s_L)).
\label{CMI}
\end{equation}
At first sight, this quantity is non-local. However, due to the short-range nature of the QCA, it turns out that Eq.~(\ref{CMI}) can be computed locally. We show this in the following, together with the equivalence to Eq.~(\ref{dS}).

We choose the size of $a_{L,R}$ to be no smaller than $2r$ so that we can make a further bipartition $a_{L,R}=a_{l,r}\sqcup a'_{l,r}$ with $|a_{l,r}|,|a'_{l,r}|\ge r$. Analogously, we consider the bipartition $s_{L,R}=s_{l,r}\sqcup s'_{l,r}$ with $|s_{l,r}|\ge r$. From the locality properties of QCAs~\cite{piroli2020quantum} we have
\begin{equation}
\rho_{a_Ls_R}= \rho_{a_l s_r}\otimes \rho_{a'_ls'_r},\;\;\;\;
\rho_{a_Rs_L}= \rho_{a_r s_l}\otimes \rho_{a'_rs'_l}.
\label{fac0}
\end{equation}
Equivalently, the above equations can be derived by using the representation of QCA in terms of matrix product unitaries, and exploiting the corresponding ``simple'' condition~\cite{cirac2017matrix,piroli2020quantum}, cf. Figs.~\ref{figS3}(b) and (c). In addition, the short-range nature of $U$ implies there should be no correlations between $s'_l$ ($s'_r$) and $a'_r$ ($a'_r$), a fact we already used in the previous section. Therefore, Eq.~(\ref{fac0}) can be further factorized as
\begin{equation}
\begin{split}
\rho_{a_Ls_R}&=\rho_{a'_l}\otimes \rho_{a_l s_r}\otimes \rho_{s'_r}, \\
\rho_{a_Rs_L}&=\rho_{a'_r}\otimes \rho_{a_r s_l}\otimes \rho_{s'_l}.
\end{split}
\label{fac}
\end{equation}
Accordingly, we have
\begin{equation}
I(a_L: s_R) = I(a_l:s_r),\;\;\;\;I(a_R: s_L) = I(a_r:s_l).
\end{equation}
As a consequence, we obtain that the CMI (\ref{CMI}) can be computed locally as
\begin{equation}
\chi=\frac{1}{2}(I(a_l: s_r) - I (a_r: s_l)).
\label{locchi}
\end{equation}
This result holds true for arbitrary R\'enyi entropies, although the corresponding mutual informations may not be positive semidefinite in general \cite{Linden2013}. 

To see that Eq.~(\ref{locchi}) is equivalent to Eq.~(\ref{dS}), we first note that it does not depend on $|\psi_{\rm prod}\rangle$ (as it can be understood from Fig.~\ref{figS3}(c)), which may even be replaced by a mixed input \footnote{This input should nevertheless be short-range correlated. Otherwise, it may mediate a nonzero correlation between $a'_{l,r}$ and $s'_{r,l}$ and thus Eq.~(\ref{fac}) may break down.}. Taking this input to be maximally mixed, we can consider $\rho_{a_ls_r}$ and $\rho_{a_rs_l}$ as the reduced states of the global CJS and $\rho_{a_{l,r}}$, $\rho_{s_{l,r}}$ should all be maximally mixed. Therefore, their entropies cancel out and the remaining entropy difference $S(\rho_{s_la_r})-S(\rho_{a_ls_r})$ exactly reproduces Eq.~(\ref{dS}).

\subsection{Equivalence between R\'enyi-2 entropy and algebra overlap}
Given two subsystems $A$ and $B$ embedded in an entire system $\Lambda$, the algebra overlap is defined as \cite{gross2012index}
\begin{equation}
\eta(\mathcal{A},\mathcal{B})=\frac{\sqrt{d_A d_B}}{d_\Lambda}\sqrt{\sum^{d_A}_{i,j=1}\sum^{d_B}_{m,n=1}|\Tr_\Lambda[e^{A^\dag}_{ij}e^B_{mn}]|^2},
\end{equation}
where $d_{A,B,\Lambda}=\dim \mathcal{H}_{A,B,\Lambda}$ and $e^{A,B}_{ij}\equiv|i^{A,B}\rangle\langle j^{A,B}|$ ($|i^{A,B}\rangle\in\mathcal{H}_{A,B}$, $i=1,2,...,d_{A,B}$). This has been used to define the index of a QCA $U$ as
\begin{equation}
{\rm ind}\equiv\log \frac{\eta(\alpha(\mathcal{A}_L),\mathcal{A}_R)}{\eta(\mathcal{A}_L,\alpha(\mathcal{A}_R))},
\end{equation}
where  $\alpha(\cdot)=U\cdot U^\dag$. Explicitly, we have
\begin{equation}
\begin{split}
&\eta(\alpha(\mathcal{A}_L),\mathcal{A}_R)=\frac{\sqrt{d_L d_R}}{d_\Lambda} \\
&\times\sqrt{\sum^{d_L}_{i,j=1}\sum^{d_R}_{m,n=1}\Tr_\Lambda[U e^{L\dag}_{ij}U^\dag e^R_{mn}]\Tr_\Lambda[U e^L_{ij}U^\dag e^{R\dag}_{mn}]},
\end{split}
\end{equation}
and $\eta(\mathcal{A}_L,\alpha(\mathcal{A}_R))$ can be written down in a similar way. Here $L$ and $R$ can be taken as small as two adjacent sites under coarse graining (such that the range of QCA is $1$). The result will not change for larger $L$ and/or $R$, which do not need to be of the same size. The double sum in the square root can be graphically represented and contracted as follows:
\begin{equation}
\begin{split}
\begin{tikzpicture}[scale=0.9]
\Text[x=-1.25,y=1]{$\sum_{i,j,m,n}$}
\Text[x=0.3,y=2.7]{$L$}
\Text[x=1,y=2.7]{$R$}
\Text[x=1.7,y=2.7]{$E$}
\draw[thick,fill=blue!10!white] (0,0) rectangle (2,0.5);
\draw[thick,fill=blue!20!white] (0,1.5) rectangle (2,2);
\Text[x=1,y=0.25]{$U^\dag$}
\Text[x=1,y=1.75]{$U$}
\draw[thick] (0.3,2) -- (0.3,2.3) (-0.3,2.3) -- (-0.3,-0.3) (0.3,0) -- (0.3,-0.3);
\draw[thick] (0.3,2.3) .. controls (0.25,2.5) and (-0.25,2.5) .. (-0.3,2.3);
\draw[thick] (0.3,-0.3) .. controls (0.25,-0.5) and (-0.25,-0.5) .. (-0.3,-0.3);
\draw[thick] (0.3,0.5) -- (0.3,0.8) (0.3,1.5) -- (0.3,1.2);
\draw[thick,fill=white] (0.3,0.8) circle (0.05) (0.3,1.2) circle (0.05);
\Text[x=0.15,y=0.8]{$i$}
\Text[x=0.15,y=1.2]{$j$}
\draw[thick] (1,0.5) -- (1,1.5) (1,2) -- (1,2.3) (1,-0.3) -- (1,0);
\draw[thick,fill=white] (1,2.3) circle (0.05) (1,-0.3) circle (0.05);
\Text[x=0.8,y=2.3]{$n$}
\Text[x=0.75,y=-0.3]{$m$}
\draw[thick] (1.7,0.5) -- (1.7,1.5);
\draw[thick] (1.7,2) -- (1.7,2.3) (2.3,2.3) -- (2.3,-0.3) (1.7,0) -- (1.7,-0.3);
\draw[thick] (1.7,2.3) .. controls (1.75,2.5) and (2.25,2.5) .. (2.3,2.3);
\draw[thick] (1.7,-0.3) .. controls (1.75,-0.5) and (2.25,-0.5) .. (2.3,-0.3);
\draw[thick,fill=blue!10!white] (3,0) rectangle (5,0.5);
\draw[thick,fill=blue!20!white] (3,1.5) rectangle (5,2);
\Text[x=4,y=0.25]{$U^\dag$}
\Text[x=4,y=1.75]{$U$}
\draw[thick] (3.3,2) -- (3.3,2.3) (2.7,2.3) -- (2.7,-0.3) (3.3,0) -- (3.3,-0.3);
\draw[thick] (3.3,2.3) .. controls (3.25,2.5) and (2.75,2.5) .. (2.7,2.3);
\draw[thick] (3.3,-0.3) .. controls (3.25,-0.5) and (2.75,-0.5) .. (2.7,-0.3);
\draw[thick] (3.3,0.5) -- (3.3,0.8) (3.3,1.5) -- (3.3,1.2);
\draw[thick,fill=white] (3.3,0.8) circle (0.05) (3.3,1.2) circle (0.05);
\Text[x=3.15,y=0.8]{$j$}
\Text[x=3.15,y=1.2]{$i$}
\draw[thick] (4,0.5) -- (4,1.5) (4,2) -- (4,2.3) (4,-0.3) -- (4,0);
\draw[thick,fill=white] (4,2.3) circle (0.05) (4,-0.3) circle (0.05);
\Text[x=3.75,y=2.3]{$m$}
\Text[x=3.8,y=-0.3]{$n$}
\draw[thick] (4.7,0.5) -- (4.7,1.5);
\draw[thick] (4.7,2) -- (4.7,2.3) (5.3,2.3) -- (5.3,-0.3) (4.7,0) -- (4.7,-0.3);
\draw[thick] (4.7,2.3) .. controls (4.75,2.5) and (5.25,2.5) .. (5.3,2.3);
\draw[thick] (4.7,-0.3) .. controls (4.75,-0.5) and (5.25,-0.5) .. (5.3,-0.3);
\end{tikzpicture}
\\
\begin{tikzpicture}[scale=0.9]
\Text[x=6.5,y=1]{$=$}
\draw[thick] (8.3,-0.85) -- (8.3,-0.15) (8.3,0.15) -- (8.3,0.85) (8.3,1.15) -- (8.3,1.85) (8.3,2.15) -- (8.3,2.85);
\draw[thick] (7.7,-0.85) -- (7.7,0.85) (7.7,1.15) -- (7.7,2.85); 
\draw[thick] (8.3,0.85) .. controls (8.25,0.9) and (7.75,0.9) .. (7.7,0.85);
\draw[thick] (8.3,-0.85) .. controls (8.25,-0.9) and (7.75,-0.9) .. (7.7,-0.85);
\draw[thick] (8.3,1.15) .. controls (8.25,1.1) and (7.75,1.1) .. (7.7,1.15);
\draw[thick] (8.3,2.85) .. controls (8.25,2.9) and (7.75,2.9) .. (7.7,2.85);
\fill[white] (7.8,-1) rectangle (7.9,3);
\draw[thick] (7.85,0.15) -- (7.85,1.85);
\draw[thick] (8.3,0.15) .. controls (8.25,0.1) and (7.9,0.1) .. (7.85,0.15);
\draw[thick] (8.3,1.85) .. controls (8.25,1.9) and (7.9,1.9) .. (7.85,1.85);
\draw[thick] (7.85,-0.15) -- (7.85,-1) (7.85,2.15) -- (7.85,3);
\draw[thick] (8.3,-0.15) .. controls (8.25,-0.1) and (7.9,-0.1) .. (7.85,-0.15);
\draw[thick] (8.3,2.15) .. controls (8.25,2.1) and (7.9,2.1) .. (7.85,2.15);
\draw[thick] (7.85,3) -- (7.55,3) -- (7.55,-1) -- (7.85,-1);
\draw[thick] (9,-0.9) -- (9,2.9) (10.5,-0.9) -- (10.5,2.9);
\draw[thick] (9,2.9) .. controls (9.1,3) and (10.4,3) .. (10.5,2.9);
\draw[thick] (9,-0.9) .. controls (9.1,-1) and (10.4,-1) .. (10.5,-0.9);
\draw[thick] (9.7,-0.85) -- (9.7,0.85) (9.7,1.15) -- (9.7,2.85);
\draw[thick] (9.7,0.85) .. controls (9.75,0.9) and (10.25,0.9) .. (10.3,0.85);
\draw[thick] (9.7,-0.85) .. controls (9.75,-0.9) and (10.25,-0.9) .. (10.3,-0.85);
\draw[thick] (10.3,-0.85) -- (10.3,0.85) (10.3,1.15) -- (10.3,2.85);
\draw[thick] (9.7,1.15) .. controls (9.75,1.1) and (10.25,1.1) .. (10.3,1.15);
\draw[thick] (9.7,2.85) .. controls (9.75,2.9) and (10.25,2.9) .. (10.3,2.85);
\draw[thick,fill=blue!10!white] (8,-0.75) rectangle (10,-0.25) (8,1.25) rectangle (10,1.75) ;
\draw[thick,fill=blue!20!white] (8,0.25) rectangle (10,0.75)  (8,2.25) rectangle (10,2.75); 
\Text[x=9,y=-0.5]{$U^\dag$}
\Text[x=9,y=0.5]{$U$}
\Text[x=9,y=1.5]{$U^\dag$}
\Text[x=9,y=2.5]{$U$}
\Text[x=11,y=1]{,}
\end{tikzpicture}
\end{split}
\label{alo}
\end{equation}
where $E=\Lambda\backslash(L\sqcup R)$. The bipartite case (i.e., $E=\emptyset$) appeared already in the original paper \cite{gross2012index} and essentially the same result has been reported in Appendix~D in Ref.~\cite{Po2016}.

It is now easy to relate the rhs of Eq.~\eqref{alo} to the R\'enyi-2 entropy of the evolution operator, see for instance~\cite{Bruno2020}. Explicitly, we first write down the CJS:
\begin{equation}
\begin{tikzpicture}[scale=0.9]
\Text[x=-0.95,y=0.25]{$|U\rangle=\frac{1}{\sqrt{d_\Lambda}}$}
\Text[x=0.3,y=1.1]{$L$}
\Text[x=1,y=1.1]{$R$}
\Text[x=1.7,y=1.1]{$E$}
\Text[x=4,y=1.1]{$L'$}
\Text[x=3.3,y=1.1]{$R'$}
\Text[x=2.6,y=1.1]{$E'$}
\draw[thick] (0.3,-0.3) -- (0.3,0.8) (1,-0.3) -- (1,0.8) (1.7,-0.3) -- (1.7,0.8);
\draw[thick] (2.6,-0.3) -- (2.6,0.8) (3.3,-0.3) -- (3.3,0.8) (4,-0.3) -- (4,0.8);
\draw[thick] (1.7,-0.3) .. controls (1.8,-0.4) and (2.5,-0.4) .. (2.6,-0.3);
\draw[thick] (1,-0.3) .. controls (1.2,-0.6) and (3.1,-0.6) .. (3.3,-0.3);
\draw[thick] (0.3,-0.3) .. controls (0.6,-0.8) and (3.7,-0.8) .. (4,-0.3);
\draw[thick,fill=blue!20!white] (0,0) rectangle (2,0.5);
\Text[x=1,y=0.25]{$U$}
\Text[x=4.5,y=0.2]{.}
\end{tikzpicture}
\end{equation}
Then it is straightforward to check that the reduced density operator on $L'R$ is nothing but half (upper or lower) of the rhs of Eq.~(\ref{alo}):
\begin{equation}
\begin{tikzpicture}[scale=0.9]
\Text[x=-3.3,y=1]{$\rho_{L'R}=\Tr_{LR'EE'}[|U\rangle\langle U|]=\frac{1}{d_\Lambda}$}
\Text[x=1,y=2.7]{$R$}
\Text[x=-0.6,y=2.7]{$L'$}
\draw[thick,fill=blue!10!white] (0,0) rectangle (2,0.5);
\draw[thick,fill=blue!20!white] (0,1.5) rectangle (2,2);
\Text[x=1,y=0.25]{$U^\dag$}
\Text[x=1,y=1.75]{$U$}
\draw[thick] (0.3,2.3) .. controls (0.25,2.5) and (-0.25,2.5) .. (-0.3,2.3);
\draw[thick] (0.3,-0.3) .. controls (0.25,-0.5) and (-0.25,-0.5) .. (-0.3,-0.3);
\draw[thick] (0.3,0.5) -- (0.3,0.7) (0.3,1.5) -- (0.3,1.3);
\draw[thick] (1,0.5) -- (1,1.5) (1,2) -- (1,2.4) (1,-0.4) -- (1,0);
\draw[thick] (1.7,0.5) -- (1.7,1.5);
\draw[thick] (1.7,2) -- (1.7,2.3) (2.3,2.3) -- (2.3,-0.3) (1.7,0) -- (1.7,-0.3);
\draw[thick] (1.7,2.3) .. controls (1.75,2.5) and (2.25,2.5) .. (2.3,2.3);
\draw[thick] (1.7,-0.3) .. controls (1.75,-0.5) and (2.25,-0.5) .. (2.3,-0.3);
\draw[thick]  (-0.6,0.7) -- (-0.6,-0.4) (-0.6,1.3) -- (-0.6,2.4);
\draw[thick] (0.3,0.7) .. controls (0.2,0.9) and (-0.5,0.9) .. (-0.6,0.7);
\draw[thick] (0.3,1.3) .. controls (0.2,1.1) and (-0.5,1.1) .. (-0.6,1.3);
\fill[white] (-0.35,0) rectangle (-0.25,2);
\draw[thick] (0.3,2) -- (0.3,2.3) (-0.3,2.3) -- (-0.3,-0.3) (0.3,0) -- (0.3,-0.3);
\Text[x=2.6,y=1]{.}
\end{tikzpicture}
\end{equation}
We thus obtain the following relation
\begin{equation}
\begin{split}
\eta(\alpha(\mathcal{A}_L),\mathcal{A}_R)^2&=d_Ld_R \Tr[\rho_{L'R}^2] \\
&=e^{S^{\max}_{L'R}-S_2(\rho_{L'R})},
\end{split}
\label{aLR}
\end{equation}
where $S^{\max}_{L'R}=\log (d_L d_R)$ is the maximal entropy of a state in $\mathcal{H}_{L'}\otimes\mathcal{H_R}$. Note that this result is consistent with $\eta\ge1$ \cite{gross2012index}. Similarly, we have
\begin{equation}
\eta(\mathcal{A}_L,\alpha(\mathcal{A}_R))^2=e^{S^{\max}_{LR'}-S_2(\rho_{LR'})}.
\label{LaR}
\end{equation}
Combining Eqs.~(\ref{aLR}) and (\ref{LaR}), we obtain
\begin{equation}
\begin{split}
{\rm ind}&\equiv\log \frac{\eta(\alpha(\mathcal{A}_L),\mathcal{A}_R)}{\eta(\mathcal{A}_L,\alpha(\mathcal{A}_R))} \\
&=\frac{1}{2}(S_2(\rho_{LR'})-S_2(\rho_{L'R})).
\end{split}
\label{ds2}
\end{equation}
Hence, we have shown that Eq.~\eqref{dS} for $\alpha=2$ is equivalent to the formulation of the index in terms of algebra overlaps.

\section{Stability against Hamiltonian evolutions} 
In this section, we provide further details on the explicit example given in the main text violating the bound~\eqref{eq:main_result}. We then prove that this represents the ``worst-case scenario'', namely the order of violation cannot be larger.

\subsection{``Worst-case scenario''}
\begin{figure}[!t]
\begin{center}
\includegraphics[width=7cm, clip]{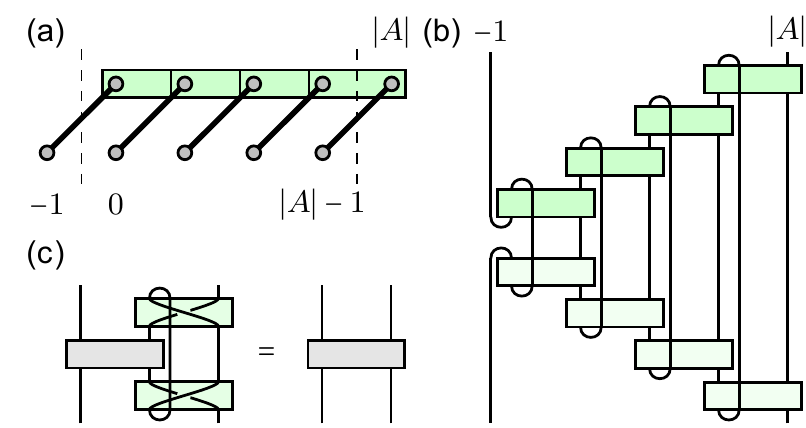}
\end{center}
   \caption{(a) CJS of $W$ (\ref{WuT}), which is the right translation evolved by a time-dependent Hamiltonian that sequentially generate two-site unitaries $u^{[j,j+1]}$ with $j=0,1,...,|A|-1$. (b) Reduced state on the $\{-1,|A|\}$, obtained by contracting the degrees of freedom in $AA'$. (c) Iterative calculation of (b) exemplified by the SWAP unitary.}
      \label{figS4}
\end{figure}

We start by rewriting the example given in the main text as
\begin{equation}
W=u^{[|A|-1,|A|]}..u^{[1,2]}u^{[0,1]}\mathbb{T},
\label{WuT}
\end{equation}
where $u^{[j,j+1]}=e^{i\theta \mathbb{S}^{[j,j+1]}}$ with $\theta= hT /|A|$, and where $\mathbb{S}^{[j,k]}$ is the SWAP operator between sites $j$ and $k$. We need to calculate the operator entanglement entropy of $W$ with respect to the bipartition $A\sqcup \bar{A}$, where $A=[0,|A|-1]$. It is immediate to see that the reduced density matrix over $A$ admits a purification in $[-1,|A|]$. In turn, this allows us to show that the operator entanglement entropy can be computed for the reduced state on $\{-1,|A|\}$, cf.~Fig.~\ref{figS4}(a). From this representation, it is immediate to see that the reduced CJS has at most rank $d^2$. This fact already implies $S_\alpha\le 2\ln d=2\ind $.
 
The reduced state $\rho_{-1,|A|}$ can be explicitly expressed as Fig.~\ref{figS4}(b)  by tracing out the degrees of freedom in $[0,|A|-1]$. Thanks to the sequential structure of the unitaries, the reduced state with the auxiliary sites traced out is actually a quantum Markov chain \cite{Hayden2004}, which inspires us to calculate $\rho_{-1,|A|}=\mathcal{E}^{|A|}(\rho_0)$ iteratively through the quantum channel (see Fig.~\ref{figS4}(c))
\begin{equation}
\begin{split}
\mathcal{E}(\rho)&=\frac{1}{d}\Tr_2[(\openone\otimes u)(\rho\otimes\openone)(\openone\otimes u^\dag)] \\
&= \sin^2\theta \rho + \cos^2\theta (\Tr_2\rho ) \otimes \frac{\openone}{d}
\end{split}
\end{equation}
starting from $\rho_0=|I_0\rangle\langle I_0|= d^{-1}\sum^d_{m,n=1}|mm\rangle\langle nn|$. One can easily check that
\begin{equation}
\mathcal{E}^n(\rho)= \sin^{2n}\theta \rho +(1-\sin^{2n}\theta) (\Tr_2\rho) \otimes \frac{\openone}{d}.
\end{equation}
When inputing $\rho_0=|I_0\rangle\langle I_0|$, we obtain
\begin{equation}
\rho_{-1,|A|}=\epsilon |I_0\rangle\langle I_0|+(1-\epsilon)\frac{\openone^{\otimes2}}{d^2}.
\end{equation}
where $\epsilon=\sin^{2|A|}\theta =\sin^{2|A|}(hT/|A|)$ is of the order of $e^{-\mathcal{O}(|A|\log |A|)}$ for large $|A|$. Since $\rho_{-1,|A|}$ (slightly) deviates from the maximally mixed state $\openone^{\otimes2}/d^2$, its R\'enyi entropy (except for $S_0$) should be smaller than $2\ln d=2\ind$, implying a violation of Eq.~(\ref{eq:main_result}). In particular, the largest violation is achieved by $S_\infty$, which turns out to be
\begin{equation}
S_\infty=-\log \|\rho_{-1,|A|}\|=2\ind - \log[1+(d^2-1)\epsilon].
\end{equation}
In addition, one can show that for $\alpha\ll \epsilon^{-1}$ the R\'enyi can be well approximated by
\begin{equation}
S_\alpha\simeq 2\ind -\frac{1}{2}\alpha(d^2-1)\epsilon^2.
\end{equation}
Therefore, the violation of the lower bound is of the order of $e^{-\mathcal{O}(|A|\log |A|)}$ for any given nonzero $\alpha$.

\subsection{General proof} 
We prove that the violation can never exceed $e^{-\mathcal{O}(|A|\log |A|)}$ in its order. Without loss of generality, we can block several adjacent sites into one such that each term $h_j(t)$ in $H(t)=\sum^N_{j=1}h_j(t)$ acts only on two adjacent (blocked) sites $j$ and $j+1$. Although the total number $N$ and the maximal local interacting strength $h\equiv\max_{j,t} \|h_j(t)\|$ should change (but still take the same order) after blocking, we use the same notations for simplicity. In this context, the range $r$ of $U_{\rm QCA}$ should be understood as the range after blocking.

To study the operator entanglement with respect to a general bipartition $\mathbb{Z}_N=[a+1,b]\sqcup [b+1,a]$, following the idea of Ref.~\cite{Osborne2006}, we decompose $H(t)$ into
\begin{equation}
H(t)= H_{[a+1,b]}(t) + H_{[b+1,a]}(t) + h_a(t) + h_b(t),
\end{equation}
where $H_{[x,y]}(t)=\sum^{y-1}_{j=x} h_j(t)$ acts nontrivially only on $[x,y]\subset\mathbb{Z}_N$. In the interaction picture, the Hamiltonian evolution can be represented as
\begin{equation}
\hat{\rm T} e^{-i\int^T_0 dt H(t)}=U_{\overline{ab}}(T)U^{({\rm I})}_{ab}(T),
\end{equation}
where 
\begin{equation}
\begin{split}
U_{\overline{ab}}(t)&=\hat{\rm T} e^{-i \int^t_0 dt'[H_{[a+1,b]}(t') + H_{[b+1,a]}(t')]},\\
U^{({\rm I})}_{ab}(t)&=\hat{\rm T} e^{-i \int^t_0 dt'[h^{({\rm I})}_a(t') + h^{({\rm I})}_b(t')]},
\end{split}
\end{equation}
with $h^{({\rm I})}_j(t)\equiv U_{\overline{ab}}(t)^\dag h_j(t)U_{\overline{ab}}(t)$. Since $U_{\overline{ab}}(t)$ is a local uniform transformation with respect to $[a+1,b]\sqcup [b+1,a]$, we can safely drop it when calculating the operator entanglement.

\begin{figure}[!t]
\begin{center}
\begin{tikzpicture}[scale=0.92]
\Text[x=-1,y=1.25,fontsize=\small]{(a)}
\draw[dashed,fill=purple!25!white] (1.75,-1.05) rectangle (5.75,1.05);
\foreach \x in {0,...,15}
{
    \draw[thick] (0.5*\x,-1) -- (0.5*\x,1);              
}
\foreach \x in {-1,...,6}
{
\fill[white] (\x+1.45,-0.6) rectangle (\x+1.55,-0.4);
}
\foreach \x in {1,...,4}
{
\fill[purple!25!white] (\x+1.45,-0.6) rectangle (\x+1.55,-0.4);
}
\foreach \x in {-1,...,6}
{    
    \draw[thick,fill=yellow!20!white] (\x+0.85,-0.6) rectangle (\x+1.65,-0.9);
    \draw[ultra thick] (\x+1,-0.6) -- (\x+1,-0.4);
    \draw[ultra thick,dotted] (\x+1.5,-0.6) -- (\x+1.5,-0.4);
}
\foreach \x in {0,...,6}
{
    \draw[thick,fill=orange!20!white] (\x+0.35,-0.1) rectangle (\x+1.15,-0.4);
}
\fill[orange!20!white] (7.35,-0.1) rectangle (7.75,-0.4);
\fill[orange!20!white] (-0.25,-0.1) rectangle (0.15,-0.4);
\draw[thick] (7.75,-0.1) -- (7.35,-0.1) -- (7.35,-0.4) -- (7.75,-0.4);
\draw[thick] (-0.25,-0.1) -- (0.15,-0.1) -- (0.15,-0.4) -- (-0.25,-0.4);
\draw[thick,fill=blue!20!white] (1.85,0.6) rectangle (5.65,0.9);
\fill[blue!20!white] (5.85,0.6) rectangle (7.75,0.9);
\fill[blue!20!white] (-0.25,0.6) rectangle (1.65,0.9);
\draw[thick] (7.75,0.6) -- (5.85,0.6) -- (5.85,0.9) -- (7.75,0.9);
\draw[thick] (-0.25,0.6) -- (1.65,0.6) -- (1.65,0.9) -- (-0.25,0.9);
\fill[white] (-0.25,0.1) rectangle (7.75,0.4);
\fill[blue!20!white] (1.35,0.1) rectangle (2.15,0.4) (5.35,0.1) rectangle (6.15,0.4);
\fill[blue!10!white] (2.15,0.1) rectangle (2.55,0.4) (1.35,0.1) rectangle (0.95,0.4) (5.35,0.1) rectangle (4.95,0.4) (6.55,0.1) rectangle (6.15,0.4);
\fill[blue!5!white] (2.95,0.1) rectangle (2.55,0.4) (0.55,0.1) rectangle (0.95,0.4) (4.55,0.1) rectangle (4.95,0.4) (6.55,0.1) rectangle (6.95,0.4);
\fill[blue!2!white] (2.95,0.1) rectangle (3.35,0.4) (0.55,0.1) rectangle (0.15,0.4) (4.55,0.1) rectangle (4.15,0.4) (7.35,0.1) rectangle (6.95,0.4);
\draw[thick] (-0.25,0.1) -- (7.75,0.1) (-0.25,0.4) -- (7.75,0.4);
\Text[x=1.75,y=-0.25,fontsize=\small]{$v_a$}
\Text[x=5.75,y=-0.25,fontsize=\small]{$v_b$}
\Text[x=1.5,y=-1.2,fontsize=\small]{$a'$}
\Text[x=2,y=1.2,fontsize=\small]{$a+1$}
\Text[x=5.5,y=-1.2,fontsize=\small]{$b'$}
\Text[x=6,y=1.2,fontsize=\small]{$b+1$}
\Text[x=-0.6,y=0.75,fontsize=\small]{$U_{\overline{ab}}$}
\Text[x=-0.6,y=0.25,fontsize=\small]{$U^{({\rm I})}_{ab}$}
\Text[x=-0.6,y=-0.5,fontsize=\small]{$U_{\rm QCA}$}
\end{tikzpicture}
\begin{tikzpicture}[scale=0.92]
\Text[x=-1,y=0.8,fontsize=\small]{(b)}
\draw[dashed,fill=purple!25!white] (1.75,-0.65) rectangle (5.75,0.65);
\foreach \x in {0,...,15}
{
    \draw[thick] (0.5*\x,-0.6) -- (0.5*\x,0.6);              
}
\fill[white] (-0.25,0.1) rectangle (3.65,0.4) (3.85,0.1) rectangle (7.75,0.4);
\fill[blue!20!white] (1.35,0.1) rectangle (2.15,0.4) (5.35,0.1) rectangle (6.15,0.4);
\fill[blue!10!white] (2.15,0.1) rectangle (2.55,0.4) (1.35,0.1) rectangle (0.95,0.4) (5.35,0.1) rectangle (4.95,0.4) (6.55,0.1) rectangle (6.15,0.4);
\fill[blue!5!white] (2.95,0.1) rectangle (2.55,0.4) (0.55,0.1) rectangle (0.95,0.4) (4.55,0.1) rectangle (4.95,0.4) (6.55,0.1) rectangle (6.95,0.4);
\fill[blue!2!white] (2.95,0.1) rectangle (3.35,0.4) (0.55,0.1) rectangle (0.15,0.4) (4.55,0.1) rectangle (4.15,0.4) (7.35,0.1) rectangle (6.95,0.4);
\draw[thick] (-0.15,0.1) rectangle (3.65,0.4) (3.85,0.1) rectangle (7.65,0.4);
\draw[ultra thick] (2,-0.6) -- (2,-0.4) (6,-0.6) -- (6,-0.4);
\fill[white] (1.45,-0.4) rectangle (1.55,-0.6);
\fill[purple!25!white]  (5.45,-0.4) rectangle (5.55,-0.6);
\draw[ultra thick,dotted] (1.5,-0.6) -- (1.5,-0.4) (5.5,-0.6) -- (5.5,-0.4);
\draw[thick,fill=orange!20!white] (1.35,-0.4) rectangle (2.15,-0.1) (5.35,-0.4) rectangle (6.15,-0.1);
\Text[x=-0.6,y=0.25,fontsize=\small]{$\tilde U^{({\rm I})}_{ab}$}
\Text[x=1.75,y=-0.25,fontsize=\small]{$v_a$}
\Text[x=5.75,y=-0.25,fontsize=\small]{$v_b$}
\Text[x=1.5,y=-0.8,fontsize=\small]{$a'$}
\Text[x=3.5,y=-0.8,fontsize=\small]{$m'$}
\Text[x=4,y=0.8,fontsize=\small]{$m+1$}
\Text[x=7.5,y=-0.8,fontsize=\small]{$n'$}
\Text[x=0,y=0.8,fontsize=\small]{$n+1$}
\Text[x=2,y=0.8,fontsize=\small]{$a+1$}
\Text[x=5.5,y=-0.8,fontsize=\small]{$b'$}
\Text[x=6,y=0.8,fontsize=\small]{$b+1$}
\end{tikzpicture}
\end{center}
   \caption{(a) Exact decomposition of $\hat{{\rm T}}e^{-i\int^T_0 dt H(t)}U_{\rm QCA}$. Only $U^{({\rm I})}_{ab}(T)$ and $v_av_b$ contribute to the operator entanglement, and the former only acts nontrivial near $a$ and $b$.  (b) Approximate unitary operator that satisfies the entanglement lower bound. Here $\tilde U^{({\rm I})}_{ab}(T)$ is factorized on $[n+1,m]\sqcup[m+1,n]$.}
      \label{figS5}
\end{figure}
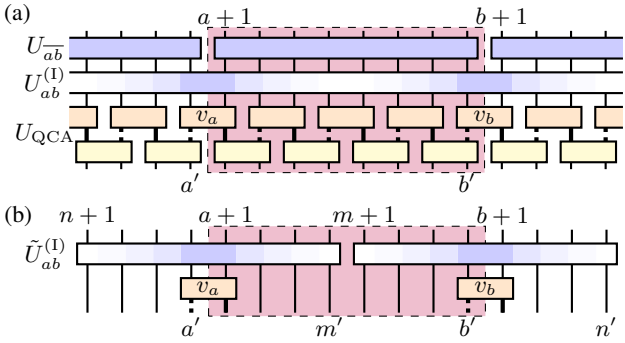

We can perform a similar simplification for the QCA part, but on the opposite side. This does not change the operator entanglement since
$(UU_{\rm LO}\otimes \mathbb{I})|I\rangle=(U\otimes U_{\rm LO}^{\rm T})|I\rangle$ 
and the transpose in the computational basis does not change the local nature. Taking a bilayer form (\ref{SMbl}) such that the $u$ layer does not entangle $[a+1,b]$ with $[b+1,a]$, we can remove all the local unitaries except for $v_a$ and $v_b$ across $a$ and $b$
, respectively. Here we have assumed $\min\{|b-a|,N-|b-a|\}\ge 2r$ so that $v_a$ and $v_b$ have no overlap. In the end, the operator entanglement entropy turns out to coincide with the entanglement entropy of (see Fig.~\ref{figS5}(a))
\begin{equation}
|\Psi\rangle=(U^{(\rm I)}_{ab}(T)v_a v_b \otimes \mathbb{I})|I\rangle=(U^{(\rm I)}_{ab}(T)\otimes v_a^{\rm T}v_b^{\rm T})|I\rangle.
\end{equation}

Intuitively, we expct $U^{({\rm I})}_{ab}(t)=\hat{\rm T} e^{-i \int^t_0 dt'[h^{({\rm I})}_a(t') + h^{({\rm I})}_b(t')]}$ to act nontrivially only near $a$ and $b$, so there should exist a good approximation 
$U^{({\rm I})}_{ab}(t)\simeq U_{[m+1,n]} U_{[n+1,m]}$,
where $U_{[x,y]}$ acts nontrivally only on $[x,y]$ and $m\in (a+1,b)$, $n\in (b+1,a)$. Again inspired by Ref.~\cite{Osborne2006}, a promising candidate can be constructed as (see Fig.~\ref{figS5}(b))
\begin{equation}
\begin{split}
\tilde U^{({\rm I})}_{ab}(t)&=\hat{\rm T} e^{-i \int^t_0 dt'[\tilde h^{({\rm I})}_a(t') + \tilde h^{({\rm I})}_b(t')]} \\
&=\hat{\rm T} e^{-i \int^t_0 dt'\tilde h^{({\rm I})}_a(t') }\hat{\rm T} e^{-i \int^t_0 dt' \tilde h^{({\rm I})}_b(t')},
\end{split}
\end{equation}
where $\tilde h^{({\rm I})}_j(t)\equiv \tilde U_{\overline{ab}}(t)^\dag h_j(t)\tilde U_{\overline{ab}}(t)$ with
\begin{equation}
\tilde U_{\overline{ab}}(t)=\hat{\rm T} e^{-i \int^t_0 dt' [H(t)-h_a(t)-h_b(t)-h_m(t)-h_n(t)]}.
\end{equation}
Since $\tilde U_{\overline{ab}}(t)$ is local with respect to $[a+1,m]\sqcup[m+1,b]\sqcup[b+1,n]\sqcup[n+1,a]$, $\tilde h^{({\rm I})}_a(t')$ ($\tilde h^{({\rm I})}_b(t')$) should be strictly supported on $[n+1,m]$ ($[m+1,n]$). Noting that
\begin{equation}
\begin{split}
U_{\overline{ab}}(t)&=\tilde U_{\overline{ab}}(t)\tilde U^{(\rm I)}_{mn}(t),\\
\tilde U^{(\rm I)}_{mn}(t)&=\hat{\rm T} e^{-i\int^t_0 dt' [\tilde h^{({\rm I})}_m(t') + \tilde h^{({\rm I})}_n(t')]},
\end{split}
\end{equation}
we can bound norm of the difference between $U^{({\rm I})}_{ab}(t)$ and $\tilde U^{({\rm I})}_{ab}(t)$ by 
\begin{equation}
\begin{split}
&\|U^{({\rm I})}_{ab}(t) - \tilde U^{({\rm I})}_{ab}(t) \| \\
\le& \int^t_0 dt' \|\tilde h^{({\rm I})}_a(t') + \tilde h^{({\rm I})}_b(t') - h^{({\rm I})}_a(t') - h^{({\rm I})}_b(t')\| \\
\le& \int^t_0 dt' \|\tilde h^{({\rm I})}_a(t') + \tilde h^{({\rm I})}_b(t') \\
&\;\;\;\;\;\;\;\;\;\;\; - \tilde U^{(\rm I)}_{mn}(t')^\dag[\tilde h^{({\rm I})}_a(t') + \tilde h^{({\rm I})}_b(t')]\tilde U^{(\rm I)}_{mn}(t')\| \\
\le& \int^t_0 dt'\int^{t'}_0 dt'' \|[ \tilde h^{({\rm I})}_a(t') + \tilde h^{({\rm I})}_b(t'),\tilde h^{({\rm I})}_m(t'') + \tilde h^{({\rm I})}_n(t'') ]\|,
\end{split}
\label{dUab}
\end{equation}
where we have used the integral versions of 
$\|e^{-ih_1}-e^{-ih_2}\|\le \|h_1-h_2\|$ and $\|e^{ih_1}h_2 e^{-ih_1}-h_2\|\le \|h_1-h_2\|$ $\forall h_{1,2}=h^\dag_{1,2}$. Recalling that $\tilde h^{({\rm I})}_x (t)$ in the interaction picture  is evolved by $H (t) -[h_a(t)+h_b(t)+h_m(t)+h_n(t)]$, which is local, we can make use of a time-dependent version of the Lieb-Robinson bound \cite{Hastings2010,Gong2020}:
\begin{equation}
\begin{split}
&\|[O_X(t_1),O_Y(t_2)]\| \le 2 \min\{|X|,|Y|\} \| O_X\| \|O_Y\| \\
&\times e^{-\kappa ({\rm dist}(X,Y)-\kappa ^{-1} \int^{t_1}_{t_2} dt\|H(t)\|_\kappa)},
\end{split}
\label{tLR}
\end{equation}
where $\kappa$ can be chosen arbitrarily as long as
\begin{equation}
\|H(t)\|_\kappa \equiv \max_{j\in \Lambda} \sum_{X\ni j} |X|\|h_X(t)\|e^{\kappa l_X}
\end{equation}
is finite for $H(t)= \sum_{X\subseteq \Lambda} h_X(t)$, which may even not be strictly local in the sense that $\|h_X(t)\|$ may decay exponentially in $l_X$, the diameter of $X$. Applying Eq.~(\ref{tLR}) to 1D and nearest-neighbor interactions, we obtain $\|H(t)\|_\kappa  \le 4h e^\kappa$ and the error bound in Eq.~(\ref{dUab}) can be estimated explicitly:   
\begin{equation}
\begin{split}
&\|U^{({\rm I})}_{ab}(T) - \tilde U^{({\rm I})}_{ab}(T) \| \\
\le&  4h^2 \left[\sum_{\substack{x=a,b,\\ y=m,n}}e^{-\kappa(|x-y|-1)} \right] 
\int^T_0 dt'\int^{t'}_0 dt'' e^{4he^\kappa(t'-t'')} \\
\le& (e^{4hTe^\kappa }-1-4hTe^\kappa )e^{-\kappa(\min_{x=a,b,y=m,n}|x-y|+1)}.
\end{split}
\end{equation}
Suppose that the size $|A|=|a-b|$ of $A=[a+1,b]$ is smaller than $N/2$, we can choose $m$ to be located at the middle of $A$ such that $\min_{x=a,b,y=m,n}|x-y|
\ge (|A|-1)/2$. In this case, we have
\begin{equation}
\|U^{({\rm I})}_{ab}(T) - \tilde U^{({\rm I})}_{ab}(T) \| \le (e^{4hTe^\kappa }-1-4hTe^\kappa )e^{-\frac{\kappa}{2}(|A|+1) }.
\end{equation}

Now let us consider another CJS
\begin{equation}
|\tilde\Psi\rangle=(\tilde U^{(\rm I)}_{ab}(T)v_av_b\otimes \mathbb{I})|I\rangle.
\end{equation}
Defining the reduced states of $|\Psi\rangle$ and $|\tilde\Psi\rangle$ on $A$ as 
$\rho_A =\Tr_{\bar A}|\Psi\rangle\langle\Psi|$ and $\tilde \rho_A = \Tr_{\bar A}|\tilde\Psi\rangle\langle\tilde\Psi|$, we have
\begin{equation}
\begin{split}
&\|\rho_A - \tilde \rho_A \| \le\frac{1}{2} \|\rho_A - \tilde \rho_A \|_1 \\
\le& \frac{1}{2}\||\Psi\rangle\langle\Psi|- |\tilde\Psi\rangle\langle\tilde\Psi|\|_1 \le \|U^{({\rm I})}_{ab}(T) - \tilde U^{({\rm I})}_{ab}(T) \| \\
\le& (e^{4hTe^\kappa }-1-4hTe^\kappa ) e^{-\frac{\kappa}{2} (|A|+1)}.
\end{split}
\end{equation}
Based on the proof in the main text, we know that $\tilde \rho_A$ satisfies the entanglement lower bound:
\begin{equation}
S_\alpha(\tilde\rho_A)\ge S_\infty(\tilde\rho_A)=-\log \| \tilde\rho_A\| \ge 2|\ind|. 
\end{equation}
Suppose that $S_\alpha(\rho_A)$ violates the lower bound, which implies
\begin{equation}
S_\infty(\rho_A)=-\log \| \rho_A\|\le S_\alpha(\rho_A) < 2|\ind|, 
\end{equation}
we have $\| \rho_A\| > e^{-2|\ind|}\ge \| \tilde\rho_A\|$. Provided that $\kappa$ is chosen such that $(e^{4hTe^\kappa }-1-4hTe^\kappa ) e^{-\frac{\kappa}{2} (|A|+1)}<e^{-2\ind}$, we can bound the difference between $S_\infty(\tilde\rho_A)$ and $S_\infty(\rho_A)$ by
\begin{equation}
\begin{split}
&|S_\infty(\tilde\rho_A)-S_\infty(\rho_A)|=\left|\log\frac{\|\rho_A\|}{\|\tilde\rho_A\|}\right| \\
\le & \frac{|\|\rho_A\|-\|\tilde\rho_A\||}{\min\{\|\rho_A\|,\|\tilde\rho_A\|\}} \le \frac{\|\rho_A - \tilde\rho_A\|}{\|\rho_A\| - \|\rho_A - \tilde\rho_A\|} \\
<&\left[ \frac{ e^{\frac{\kappa}{2}(|A|+1)-2|\ind|}}{e^{4hTe^\kappa }-1-4hTe^\kappa } -1\right]^{-1},
\end{split}
\end{equation}
where we have used $\log(1+x)\le x$, $\forall x\in(-1,\infty)$ and the triangle inequality of the operator norm. Accordingly, we can lower bound $S_\alpha(\rho_A)$ by
\begin{equation}
\begin{split}
S_\alpha(\rho_A)&\ge S_\infty(\rho_A)= S_\infty(\tilde\rho_A)- |S_\infty(\tilde\rho_A)-S_\infty(\rho_A)| \\
&> 2|\ind| -\left[ \frac{ e^{\frac{\kappa}{2}(|A|+1)-2|\ind|}}{e^{4hTe^\kappa }-1-4hTe^\kappa } -1\right]^{-1}. 
\end{split}
\label{vb}
\end{equation}
If $e^{\frac{\kappa}{2}(|A|+1)-2|\ind|}>e^{4hTe^\kappa}$, we may drop $1+4hTe^\kappa $ in the denominator and optimize the lower bound to be
\begin{equation}
S_\alpha(\rho_A)> 
2|\ind| - 
\left[e^{-2|\ind| }\left(\frac{|A|+1}{8ehT}\right)^{\frac{|A|+1}{2}} -1\right]^{-1}.
\label{lb}
\end{equation}
Here $\kappa$ is chosen such that $4hT e^\kappa=(|A|+1)/2$, which follows from the fact that $x^\beta e^{-x}$ ($\beta>0$) reaches its maximal $(\beta/e)^\beta$ at $x=\beta$. A sufficient self-consistent condition for the validity of Eq.~(\ref{lb}) could be $|A|>8\max\{e^2hT,|\ind|\}-1$. 
This result (\ref{lb}) implies that, given $h$, $T$ and $\ind$, the possible violation is superexponentially suppressed by $|A|$ as $e^{-\mathcal{O}(|A|\log |A|)}$ for large $|A|$.

As an application, we consider $U_{\rm QCA}=\mathbb{T}$ and time-independent $H=\sum^N_{j=1} h_j$, which can be in the deep MBL phase \cite{Po2016}, 
and the time evolution of $U=e^{-iHT}U_{\rm QCA}$ for $n$ periods. We can rewrite the total time-evolution operator into
\begin{equation}
(e^{-iHT}\mathbb{T} )^n = \overleftarrow{\prod^n_{j=1} }e^{-i\mathbb{T}^{n-j}H\mathbb{T}^{j-n} T}\mathbb{T}^n. 
\end{equation}
which takes the form of Eq.~(\ref{UH}) 
with $U_{\rm QCA}=\mathbb{T}^n$ and
$h_j(t)= h_{j\pm (n-s)},\;(s-1)T<t< sT$
for $s=1,2,...,n$. Therefore, Eq.~(\ref{lb}) holds true for $h= \max_j\| h_j\|$ and $T\to nT$, $|\ind | \to n|\ind |$. It follows that, due to the negligible violation, the operator entanglement grows linearly up to $n\sim \mathcal{O}(|A|)$, ruling out the possibility of MBL.

\end{document}